\documentclass[10pt,journal,letterpaper,twocolumn,twoside,nofonttune]{IEEEtran}
\usepackage{mathpazo}
\usepackage{times}

\IEEEoverridecommandlockouts
\usepackage{amssymb}
\usepackage{amsmath}
\usepackage{graphicx}
\usepackage{stmaryrd}       
\usepackage{subfigure}
\usepackage[ruled,vlined]{algorithm2e}

\usepackage{cite}     
 
\usepackage{upref}

\title{\Huge How to Construct Polar Codes\\[0.20ex]}

\author{\large%
Ido Tal,\,\,\IEEEmembership{Member,~IEEE},
and 
Alexander Vardy,\,\,\IEEEmembership{Fellow,~IEEE}
\vspace{-2.250ex}
\thanks{%
    Submitted for publication October 30, 2011.
    Revised \today.
    The material in this paper was presented in part
    at the Workshop~on~Information Theory,
    Dublin, Ireland, August 2010.
}
\thanks{Ido Tal is with the
    Department of Electrical~Engineering,
    Technion --- Israel Institute of Technology,
    Haifa, 32000, Israel
    (e-mail: \texttt{idotal@ieee.org}).}
\thanks{Alexander Vardy is with the
    Department of Electrical and Computer Engine\-ering
    and the Department of Computer Science and Engineering,
    University of California San Diego,
    La Jolla, CA 92093--0407, U.S.A.
   (e-mail: \texttt{avardy@ucsd.edu}).}
}

\renewcommand{\markboth}[2]
{\renewcommand{\leftmark}{#1}\renewcommand{\rightmark}{#2}}

\newcommand{\cA}{{\cal A}}

\newcommand{\cP}{{\cal P}}

\DeclareMathAlphabet{\mathbfsl}{OT1}{ppl}{b}{it} 

\newcommand{\uuu}{\mathbfsl{u}} 
\newcommand{\vvv}{\mathbfsl{v}}

\newcommand{\yyy}{\mathbfsl{y}}

\newcommand{\be}[1]{\begin{equation}\label{#1}}
\newcommand{\ee}{\end{equation}} 

\newcommand{\Eq}[1]{(\ref{#1})}

\renewcommand{\le}{\leqslant} 
\renewcommand{\leq}{\leqslant}
\renewcommand{\ge}{\geqslant} 
\renewcommand{\geq}{\geqslant}

\DeclareMathOperator{\argmax}{argmax}

\newcommand{\twobibs}[2]{#2} 

\newcommand{\Arikan}{Ar\i kan}

\newcommand{\Prob}{\mathrm{Prob}}

\newcommand{\floor}[1]{\left\lfloor #1 \right\rfloor}
\newcommand{\calA}{\mathcal{A}}
\newcommand{\calX}{\mathcal{X}}
\newcommand{\calY}{\mathcal{Y}}
\newcommand{\calZ}{\mathcal{Z}}

\newcommand{\Bernoulli}{\mathrm{Bernoulli}}

\newcommand{\bary}{\bar{y}}
\newcommand{\yerasure}{y_{\scriptscriptstyle{?}}}
\newcommand{\baryerasure}{\bary_{\scriptscriptstyle{?}}}
\newcommand{\barz}{\bar{z}}
\newcommand{\zmerged}{z_{1,2}}
\newcommand{\deltaI}{\mathrm{deltaI}}

\newcommand{\w}{w}
\newcommand{\barw}{\bar{\w}}

\newcommand{\barzmerged}{\barz_{1,2}}

\newcommand{\PFER}{P_{\Wunderlying,n}(k)}

\newcommand{\errorProb}{P_e}
\newcommand{\errorProbUpperBound}{\overline{P_e}}

\newcommand{\BSC}{\mathrm{BSC}}

\newcommand{\Wunderlying}{\mathbb{W}}
\newcommand{\Wbit}{\mathcal{W}}
\newcommand{\Wgeneric}{\Wbit}

\newcommand{\inputPairsCount}{L}
\newcommand{\outputPairsCount}{\nu}
\newcommand{\Wupgraded}{\mathcal{Q}'}
\newcommand{\Wdegraded}{\mathcal{Q}}
\newcommand{\intermediateChannel}{\cP}
\newcommand{\Wequivalent}{\Wgeneric'}
\newcommand{\WupgradedTriple}{\Wupgraded_{123}}

\newcommand{\WupgradedDoubleShifted}{\Wupgraded_{23}}
\newcommand{\calYupgradedTriple}{\calYupgraded_{123}}

\newcommand{\calYupgradedDoubleShifted}{\calYupgraded_{23}}
\newcommand{\zTriple}{z_{123}}

\newcommand{\zDoubleShifted}{z_{23}}

\newcommand{\Wgood}{\Wgeneric_{\varoast}}
\newcommand{\Wbad}{\Wgeneric_{\boxast}}
\newcommand{\Qgood}{\Wdegraded_{\varoast}}
\newcommand{\Qbad}{\Wdegraded_{\boxast}}

\newcommand{\reals}{\mathbb{R}}

\newcommand{\contLR}{\theta}

\newcommand{\calYdegraded}{\calZ}
\newcommand{\calYupgraded}{\calZ'}
\newcommand{\ydegraded}{z}
\newcommand{\yupgraded}{z'}

\newcommand{\Rexact}{R_\mathrm{exact}}
\newcommand{\Rdegraded}{R_\mathrm{degraded}}
\newcommand{\Rupgraded}{R_\mathrm{upgraded}}

\newcommand{\xor}{\oplus}

\newcommand{\degraded}{\preccurlyeq}
\newcommand{\upgraded}{\succcurlyeq}
\newcommand{\equivalent}{\equiv}

\newcommand{\mysett}[1]{\{#1\}}
\newcommand{\myset}[1]{\left\{#1\right\}}
\newcommand{\sizee}[1]{|#1|}
\newcommand{\size}[1]{\left|#1\right|}
\newcommand{\LR}{\mathrm{LR}}
\newcommand{\errorBlock}{e_\mathrm{Block}}

\newcommand{\BMS}{BMS}

\SetAlFnt{\small}

\SetKwFunction{degradingMerge}{degrading\_merge}
\SetKwFunction{upgradingMerge}{upgrading\_merge}
\SetKwFunction{insertRightmost}{insertRightmost}
\SetKwFunction{getMin}{getMin}
\SetKwFunction{removeMin}{removeMin}
\SetKwFunction{valueUpdated}{valueUpdated}

\SetKwInOut{Input}{input}\SetKwInOut{Output}{output}
\newcommand{\WAlg}{\mathcal{W}}
\newcommand{\QAlg}{\mathcal{Q}}
\SetKwData{ZAlg}{Z}
\newcommand{\QprimeAlg}{\mathcal{Q}'}
\SetKwData{datum}{d}
\SetKwData{datumLeft}{dLeft}
\SetKwData{datumRight}{dRight}
\SetKwData{leftAlg}{left}
\SetKwData{rightAlg}{right}
\SetKwData{nullAlg}{null}

\SetKwFunction{calcDeltaI}{calcDeltaI}

\DontPrintSemicolon
\SetAlgoSkip{-5pt}

\newcommand{\binaryRep}[1]{\langle #1 \rangle_2}

\newtheorem{lemm}{Lemma}
\newtheorem{theo}[lemm]{Theorem}
\newtheorem{prop}[lemm]{Proposition}
\newtheorem{coro}[lemm]{Corollary}

\newcommand{\Gm}{G^{\raisebox{1pt}{${\scriptscriptstyle\otimes}\scriptstyle m$}}} 
\newcommand{\ATgood}{\mathop{\raisebox{0.25ex}{\footnotesize$\varoast$}}}
\newcommand{\ATbad}{\mathop{\raisebox{0.25ex}{\footnotesize$\boxast$}}}
\newcommand{\deff}{\mbox{$\stackrel{\rm def}{=}$}}

\begin{document}

\maketitle

\begin{abstract}
\looseness=-1
A method for efficiently constructing polar codes~is presented and
analyzed. Although polar codes are explicitly defined, 
straightforward construction is intractable since the result\-ing 
polar bit-channels 
have an output alphabet that grows
exponentially with the code length. 
Thus the core problem that~needs to be solved is that of
faithfully approximating a~bit-channel~with an intractably
large alphabet by another channel having a manageable
alphabet size. We devise two approximation methods which
``sandwich'' the original bit-channel between a degraded
and an upgraded version thereof. Both approximations 
can be efficiently computed, and turn out to be extremely 
close in practice. We also\linebreak provide theoretical analysis
of our construction algorithms,~prov-ing 
that for any fixed $\varepsilon > 0$~and all
sufficiently large code~lengths~$n$, polar codes whose
rate~is~within $\varepsilon$ of channel capacity 
can~be~con-structed in time and space that are both
linear in $n$.
\end{abstract}

\begin{keywords} 
channel polarization, 
channel degrading and upgrading,
construction algorithms,
polar codes
\vspace{-1.00ex}
\end{keywords}

\section{Introduction}
\label{sec:introduction}

\noindent 
\PARstart{P}{olar} codes, invented by \Arikan~\cite{Arikan:09p},
achieve the capacity of arbitrary binary-input symmetric DMCs.
Moreover, they have low encoding and decoding complexity and
an~exp\-li\-cit construction. 
%
Following \Arikan's seminal paper~\cite{Arikan:09p},
his results have been extended in a~variety of important ways.
In \cite{STA:09a}, polar codes have been generalized to symmetric
DMCs with \emph{non-binary} input alphabet.
In \cite{KSU:10p}, the polarization phenomenon has been
studied for \emph{arbitrary kernel matrices}, rather
than \Arikan's original $2 \times 2$ polarization kernel,
and error~exponents were derived for each such kernel. 
It was shown~in~\cite{TV:11} that, under \emph{list-decoding},
polar codes can achieve remarkably good performance at short 
code lengths.
In terms of applications, polar coding has been used with great
success in the context of 
multiple-access channels \cite{STY:10a,AbbeTelatar:10a},
wiretap channels~\cite{MahdavifarVardy:11p},
data compression~\cite{Arikan:10a,Abbe:11a}, 
write-once channels~\cite{BurshteinStrugatski:12a},~and
channels with memory~\cite{Sasoglu:11c}.
In this paper, however, we will~restrict~our attention
to the original setting introduced by \Arikan\ in~\cite{Arikan:09p}. 
Namely, we focus on binary-input, discrete, memoryless, 
symmetric~channels, 
with the standard $2 \,{\times}\, 2$ polarization~kernel~under 
standard successive cancellation decoding. 

Although the construction of polar codes is \emph{explicit}, there is
only one known instance --- namely, the binary erasure~channel (BEC) ---
where the
construction is also \emph{efficient}. A first attempt at an efficient
construction of polar codes in the general case was made by Mori and
Tanaka~\cite{Mori:10z,MoriTanaka:09c}. 
Specifically, it is shown in~\cite{Mori:10z} that
a key step in the construction of polar bit-channels
can~be viewed as an instance of density 
evolution~\cite{RichardsonUrbanke:08b}. 
Based on this observation, Mori and Tanaka~\cite{MoriTanaka:09c}
proposed a construction algorithm utilizing convolutions, 
and proved that the number of convolutions needed scales
linearly with the code length.
However, as indeed noted in \cite{Mori:10z}, it is
not clear how one would implement such convolutions to be
sufficiently precise on one hand while being tractable 
on the other hand.

\looseness=-1
In this paper, we further extend the ideas of
\cite{Mori:10z,MoriTanaka:09c}. 
An~exact implementation of the convolutions discussed
in \cite{Mori:10z,MoriTanaka:09c}\linebreak
implies an algorithm with memory requirements
that grow exponentially with the code length. It is thus impractical. 
Alternatively, one could use quantization (binning) to try and reduce 
the memory requirements. However, for such quantization scheme 
to be of interest, it must satisfy two conditions. 
First, it must be fast enough, which usually translates 
into a rather small number of quantization levels (bins). 
Second, after the calculations have been carried out, we must
be able to interpret them in a precise manner. That is, the
quantization operation introduces inherent inaccuracy into
the computation, which we should be able to account for
so as to ultimately make a precise statement.

Our aim in this paper is to provide a method by which~polar codes 
can be efficiently constructed. Our main contribution~consists of
two approximation methods. In both methods, the memory limitations are
specified, and not exceeded. One method is used to get a lower bound
on the probability of error of each polar bit-channel while the other 
is used to obtain an upper bound. The quantization used to derive a lower 
bound on the probability of error is called a \emph{degrading quantization},
while the other is called an \emph{upgrading quantization}. Both
quantizations transform the ``current channel'' into a new one with a
smaller output alphabet. The degrading quantization results in a
channel degraded with respect to the original one, while the upgrading
quantization results in a channel such that the original channel is
degraded with respect to it. 

The fidelity of both degrading and upgrading approximations
is a function of a parameter $\mu$,
which can be freely set to~an arbitrary integer value.
Generally speaking, the larger $\mu$ is the better the approximation. 
The running time needed in order to approximate all $n$ polar bit-channels 
is $O(n \cdot \mu^2 \log \mu)$.

Our results 
relate to both theory and practice of polar codes.
In practice, it turns out that the degrading~and~upgrading approximations
are typically very close,
even for relatively small values of the fidelity parameter $\mu$.
This is illustrated in what follows with the help of
two examples.

\begin{figure*}
\centering
\includegraphics[width=122mm]{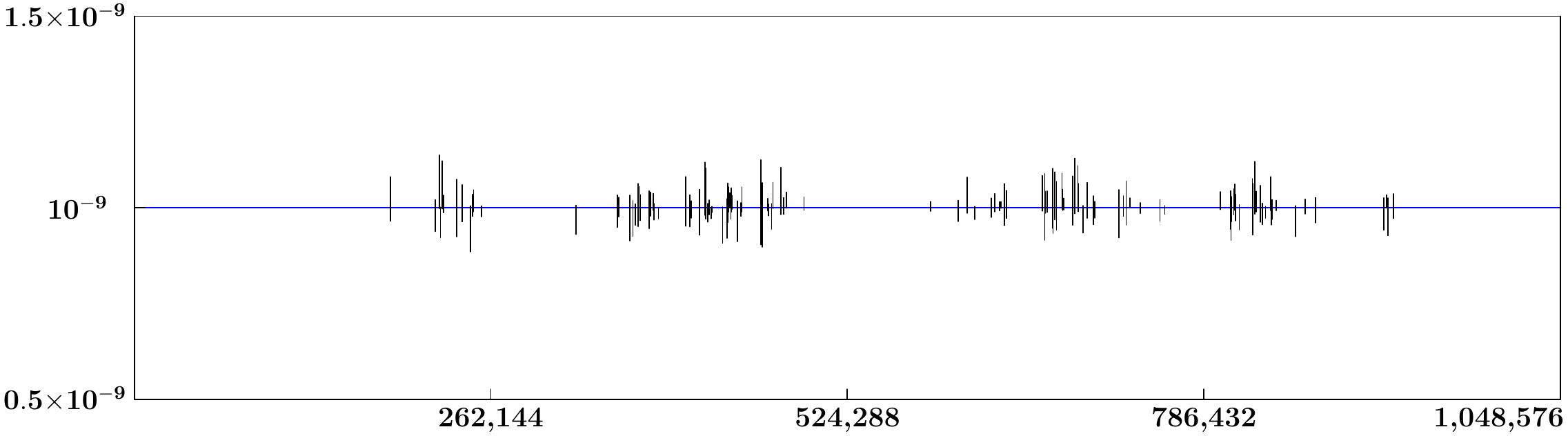}
\caption{%
\hspace*{-1ex}Upper and lower bounds on 
the bit-channel probabilities of error 
for a polar code of length $n = 1,048,576$ on BSC($0.11$),
computed using degrading and upgrading algorithms with $\mu = 256$.
Only those $132$ bit-channels for which the gap between 
the upper and lower bounds crosses the $10^{-9}$ threshold 
are shown.
}
\label{fig:BSCthreshold}
\end{figure*}

\begin{figure*}
\centering
$\,$\\[2.00ex]
\subfigure[Binary symmetric channel $\mathrm{BSC}(0.001)$]%
{\label{subfig:011}\includegraphics[width=60mm]{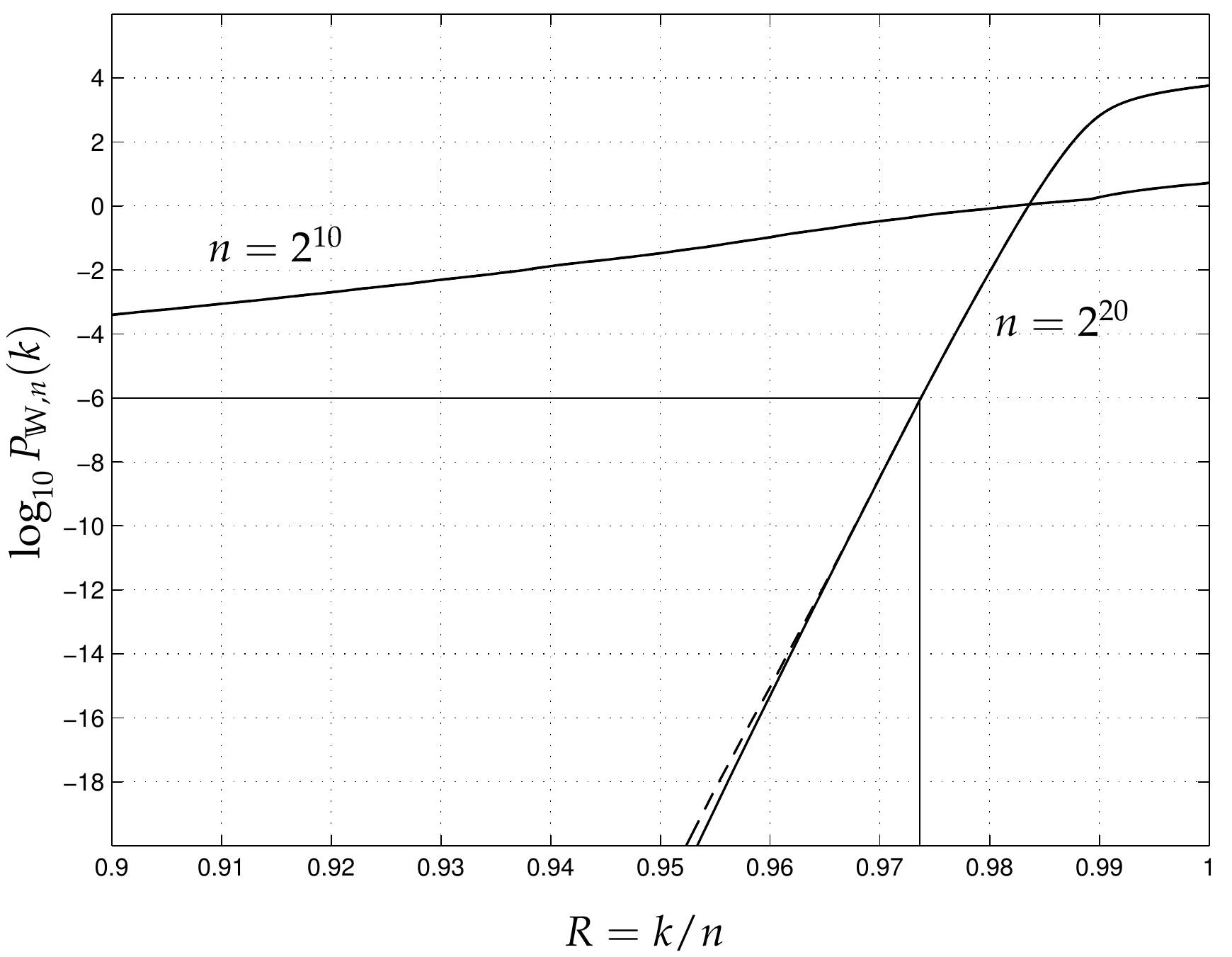}}
\hspace*{9ex}
\subfigure[binary-input AWGN channel with 
$E_s/N_0=5.00$\,dB]%
{\label{subfig:0001}\includegraphics[width=60mm]{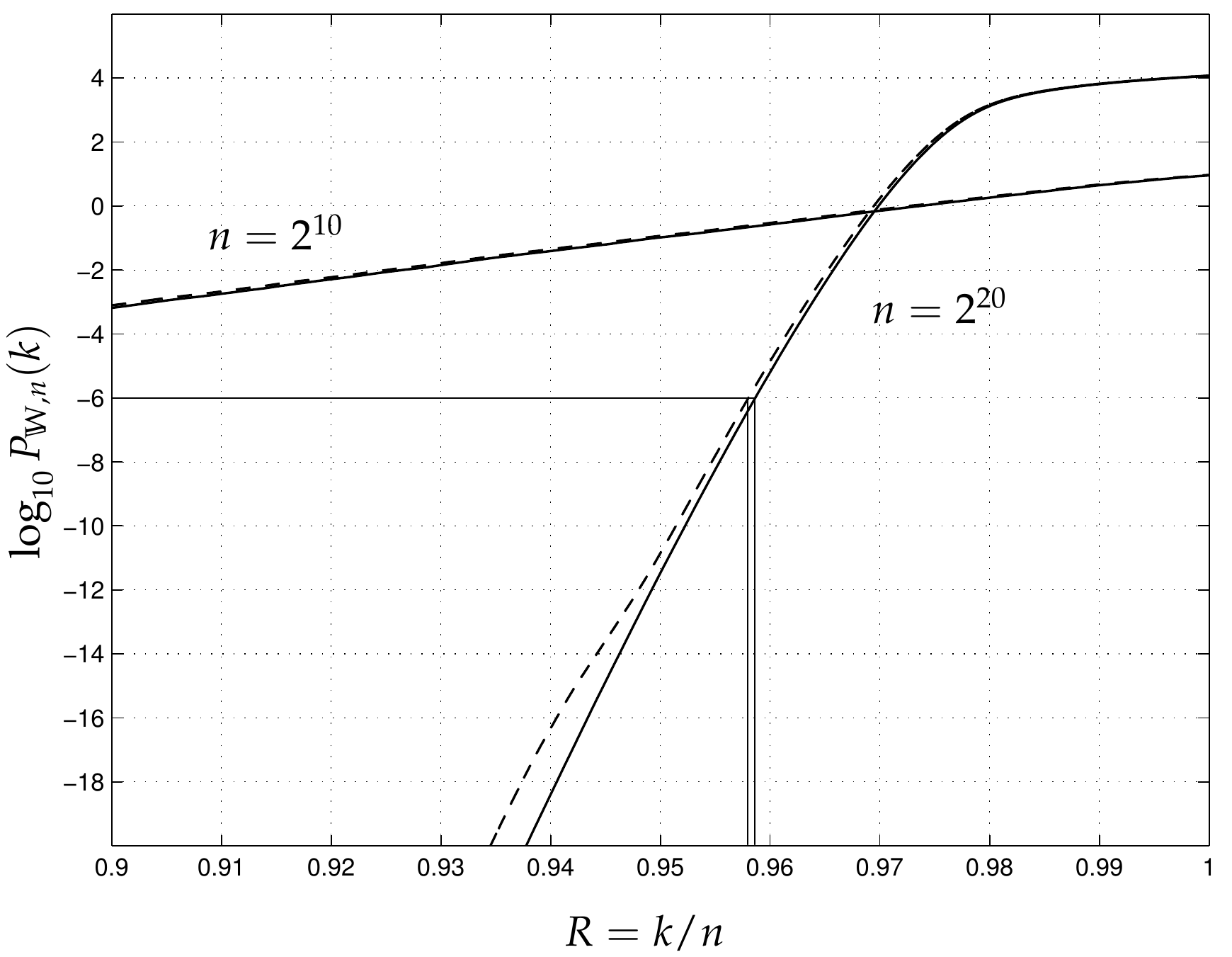}}
\caption{%
\hspace*{-1ex}Upper and lower bounds on $P_{\Wunderlying,n}(k)$
as a function of rate $R=k/n$, for two underlying channels
and two code lengths $n = 2^{10}$ and $n = 2^{20}$.
The upper bound is dashed while the lower bound is solid. 
For both channels, the difference between the bounds can only 
be discerned in the plot corresponding to $n=2^{20}$.%
}
\label{fig:BSCplots}
\end{figure*}

\vspace{0.50ex}
\noindent
{\bf Example\,1.} 
Consider a polar code of length $n = 2^{20}$ for the binary symmetric
channel (BSC) with crossover probability~$0.11$. Let 
$\Wbit_0,\Wbit_1,\ldots,\Wbit_{n-1}$ be the corresponding bit-channels
(see the next section for a rigorous definition of a bit-channel).
The basic task in the construction of polar codes is that of classifying
bit-channels into those that are ``good'' and those that are ``bad.''
Let $\errorProb(\Wbit_i)$ denote the probability of error on the 
$i$-th bit-chan-nel (see \Eq{eq:errorProbDef} for a precise definition
of this quantity)~for~$i = 0,1,\ldots,n\,{-}\,1$.
We arbitrarily choose a threshold of $10^{-9}$
and say that the $i$-th bit channel is good if 
$\errorProb(\Wbit_i) \leq 10^{-9}$ and bad otherwise.
How well do our algorithms perform~in~determining for
each of the $n$ bit-channels whether it is good or bad?

Let us set $\mu = 256$ and compute upper and lower bounds 
on $\errorProb(\Wbit_i)$ for all $i$, using the
degrading and upgrading quantizations, respectively.
The results of this computation are illustrated in 
Figure\,\ref{fig:BSCthreshold}.
In $1,048,444$ out of the $1,048,576$ cases, 
we~can provably 
classify the bit-channels into good and bad. 
Figure\,\ref{fig:BSCthreshold} depicts
the remaining $132$ bit-channels 
for which the upper
bound is above the threshold whereas the lower bound
is below the threshold. 
The horizontal axis in 
Figure\,\ref{fig:BSCthreshold}
is the bit-channel index while the vertical
axis is the gap between the two bounds. 
We see that the gap between the upper and lower bounds, 
and thus the remaining uncertainty as to the true 
value of $\errorProb(\Wbit_i)$, is very small in all cases.
\hfill\raisebox{-0.50ex}{$\Box$}

\vspace{1.00ex}
\noindent
{\bf Example\,2.} 
Now suppose we wish to construct a polar code~of a given length $n$
having the \emph{best possible rate} while guaranteeing a certain
block-error probability $P_{\rm block}$ under successive cancellation
decoding. \Arikan~\cite[Proposition 2]{Arikan:09p} provides\footnote
{%
In \cite{Arikan:09p}, \Arikan\ uses the Bhattacharyya parameter 
$Z(\Wbit_i)$ instead of the probability of error $\errorProb(\Wbit_i)$.
As we shall see shortly, this is of no real importance.
} 
a union bound on the block-error rate of polar codes:
\be{union}
P_{\rm block}
\ \le \
\sum_{i \in \cA} \errorProb(\Wbit_i)
\ee
where $\cA$ is the \emph{information set} for the code
(the set of unfrozen bit-channels). The construction
problem for polar codes can be phrased (cf.~\cite[Section\,IX]{Arikan:09p})
as the problem 
of choosing an information set $\cA$ of a given size
$|\cA| = k$ so as to minimize the right-hand side of~\Eq{union}.
Assuming the underlying channel $\Wunderlying$ and
the code length $n$ are fixed, let
\be{PWnk-def}
P_{\Wunderlying,n}(k)
\ \ \deff \ \
\min_{|\cA| = k} \, \sum_{i \in \cA} \!\errorProb(\Wbit_i)
\ee
Using our degrading and upgrading algorithms, we 
can~effici\-ent\-ly compute upper and lower bounds on $P_{\Wunderlying,n}(k)$.
These~are plotted in Figure\,\ref{fig:BSCplots} for two
underlying channels: 
BSC~with cross\-over probability $0.001$ 
and the binary-input AWGN channel with a symbol SNR of $5.00$\,dB
(noise variance $\sigma^2=0.1581$).
In all\footnote{%
The initial degrading (upgrading) transformation of the
binary-input~con\-tinous-output AWGN channel to a binary-input 
channel with a finite output alphabet was done according to the method    
of~Section~\ref{sec:continuousChannels}. For that calculation,
we used a finer value of $\mu = 2000$. Note that the initial
degrading (upgrading) transformation is performed only once.
}
our calculations, the value of $\mu$ did not exceed $512$.

\looseness=-1
As can be seen from Figure\,\ref{fig:BSCplots},
the bounds effectively coincide.
As an example, consider polar codes of length $2^{20}$
and suppose we wish to guarantee
$P_{\Wunderlying,n}(k) \le 10^{-6}$.
What is the best possible rate of such a code?
According~to~Figure\,\ref{subfig:011}, we can
efficiently construct (specify the rows of 
a~generator matrix) a polar 
code of rate $R = 0.9732$.
On the other hand, we can also prove that there is 
no choice of an information set $\cA$ in
\Eq{PWnk-def} that would possibly produce
a polar code of rate $R \:{\ge}\: 0.9737$.
According~to~Figure\,\ref{subfig:0001}, 
the corresp\-onding numbers for the binary-input
AWGN channel are $0.9580$ and $0.9587$.
In practice, such minute differences in
the code rate are 
negligible.
\hfill\raisebox{-0.50ex}{$\Box$}

\vspace{1.00ex}
From a theoretical standpoint, one of our main contributions is the
following theorem. In essence, the theorem asserts that 
capacity-achieving polar codes can be constructed in
time that is polynomial (in fact, linear) in their length $n$.\vspace{0.50ex}

\begin{theo}
\label{theo:polyConstruction}
Let $\Wunderlying$ be a
binary-input, symmetric, discrete me\-moryless channel of 
capacity $I(\Wunderlying)$. Fix arbitrary real
constants $\varepsilon > 0$ and $\beta < 1/2$. Then 
there exists an even integer 
\be{mu0-def}
\mu_0 \ = \ \mu_0(\Wunderlying,\varepsilon,\beta)\ ,
\ee
which does \emph{not} depend on the code length $n$, 
such that the following holds. 
For all even integers $\mu \geq \mu_0$ and 
all sufficiently large code lengths $n=2^m$, 
there is a construction algorithm with running 
time $O(n \cdot \mu^2 \log \mu)$ that produces
a polar code for $\Wunderlying$ of rate
$
R \ge I(\Wunderlying) - \varepsilon
$\,
such that
$
P_{\rm block} \le \smash{2^{-n^\beta}}
$,
where $P_{\rm block}$ is the probability 
of codeword error under successive~cancellation decoding.\vspace{1.00ex}
\end{theo}

We defer the proof of Theorem\,\ref{theo:polyConstruction} to
Section~\ref{sec:analysis}. Here, let us briefly discuss two
immediate consequences of this theorem. First, observe that
for a given channel $\Wunderlying$ and any fixed $\varepsilon$
and $\beta$, the integer $\mu_0$ in \Eq{mu0-def} is a constant.
Setting our fidelity parameter 
in Theorem\,\ref{theo:polyConstruction}
to $\mu = \mu_0$ thus yields a construction algorithm 
with running time that is \emph{linear in $n$}.
Still, some might argue that the complexity of construction
in Theorem\,\ref{theo:polyConstruction} does depend on 
a fidelity parameter $\mu$, and this is unsatisfactory.
The following corollary eliminates this dependence altogether,
at the expense of super-linear construction 
complexity.\vspace{0.50ex}

\begin{coro}
Let $\Wunderlying$ be a
binary-input, symmetric, discrete me\-moryless channel of 
capacity $I(\Wunderlying)$. Fix arbitrary real
constants $\varepsilon > 0$ and $\beta < 1/2$. Then
there is a construction algorithm with running time 
$O(n \log^2\!n \log\log n)$
that for all sufficiently large code lengths $n$, 
produces a polar code for $\Wunderlying$ of rate
$R \ge I(\Wunderlying) - \varepsilon$\,
such that
$
P_{\rm block} \le \smash{2^{-n^\beta}}
$.
\end{coro}

\begin{proof}
Set $\mu=2\floor{\log_2 n}$ in Theorem~\ref{theo:polyConstruction}
(in fact, we could have used any function of $n$ that grows
without bound).\vspace{0.50ex}
\end{proof}

\looseness=-1
We would now like to draw the reader's attention to 
what~The\-orem~\ref{theo:polyConstruction} \emph{does not} assert. 
Namely, given $\Wunderlying$, $\varepsilon$ and $\beta$, the theorem 
does not tell us how large $n$ must be, only that some values~of $n$ 
are large enough.
In fact, given $\Wunderlying$, $\varepsilon,\beta$, 
how large does $n$ need to be in order to guarantee the
\emph{existence} of a polar code with
$R \ge I(\Wunderlying) - \varepsilon$
and $P_{\rm block} \le \smash{2^{-n^\beta}}$,
let alone the complexity of its construction?
This is one of the central questions in the
theory of polar codes. Certain lower bounds
on this value of $n$ are given in
\cite{GoliHassaniUrbanke:12c}.
In the other direction, the exciting recent
result of Guruswami and Xia~\cite[Theorem\,1]{GuruswamiXia:13aa}
shows that for any fixed $W$ and $\beta \le 0.49$, this
value of $n$ grows as a polynomial in $1/\varepsilon$.
The work of \cite{GuruswamiXia:13aa} further shows
that, for any fixed $W$ and $\beta$, the 
parameter $\mu_0$ in \Eq{mu0-def} can be also taken
as a polynomial in $1/\varepsilon$.
 
\looseness=-1
The rest of this paper is oragnized as follows. In
Section~\ref{sec:polarCodes}, we briefly review polar codes and 
set up the necessary notation. 
Section~\ref{sec:degradeUpgrade} is devoted to 
channel degrading and upgrading relations, 
that will be important for us later on.
In Section~\ref{sec:highLevelAlg}, we give a high
level description of our algorithms for approximating 
polar bit-channels. The
missing details in Section~\ref{sec:highLevelAlg} are then fully
specified in Section~\ref{sec:mergingFunctions}. Namely, we show 
how to reduce the output alphabet of a channel so as 
to get either~a~de\-graded or an upgraded version 
thereof.
In Section~\ref{sec:continuousChannels}, we show how to either
degrade or upgrade a channel with continuous~output into 
a~channel with a finite output alphabet of specified~size. In
Section~\ref{sec:variationsOnTheme}, we discuss certain 
improvements to our general algorithms 
for a specialized case. The accuracy of the 
(improved) algorithms is then analyzed in Section~\ref{sec:analysis}.

\vspace{2ex}
\section{Polar Codes}
\label{sec:polarCodes}

In this section we briefly review polar codes with the primary aim of
setting up the relevant notation. We also indicate where the difficulty 
of constructing polar codes lies.

Let $\Wunderlying$ be the underlying memoryless channel through
which we are to transmit information. 
If the input alphabet of $\Wunderlying$ is $\calX$ and its 
output alphabet is $\calY$, we write $\Wunderlying\!: \calX \to \calY$. 
The probability of observing $y \in \calY$ given that $x \in\calX$ 
was transmitted is denoted by $\Wunderlying(y|x)$. 
We assume throughout that $\Wunderlying$ has binary 
input and so $\calX=\mysett{0,1}$. We also assume that
$\Wunderlying$ is symmetric.
As noted in~\cite{Arikan:09p}, a binary-input channel~$\Wunderlying$
is symmetric if and only if there exists a permutation 
$\pi$ of $\calY$
such that $\pi^{-1} = \pi$ (that is, $\pi$ is an involution) and
$\Wunderlying(y|1) = \Wunderlying(\pi(y)|0)$ 
for all $y \in \calY$ (see \cite[p.\,94]{Gallager:68b} 
for an equivalent definition). 
When~the\linebreak
permutation is understood from the context, we abbreviate $\pi(y)$ as
$\bary$, and say that $\bary$ and $y$ are \emph{conjugates}. For now, we
will further assume that the output alphabet $\calY$ of $\Wunderlying$
is finite. This assumption will be justified in 
Section~\ref{sec:continuousChannels}, where 
we show how to 
deal with channels that have continuous output.

Denote the length of the codewords we will be transmitting over $\Wunderlying$ by $n = 2^m$. Given $\yyy = (y_0,y_1,\ldots,y_{n-1}) \in \calY^n$ 
and $\uuu = (u_0,u_1,\ldots,u_{n-1}) \in \calX^n$, 
let 
$$
\Wunderlying^n(\yyy|\uuu)
\ \ \deff\ \
\prod_{i=0}^{n-1} \Wunderlying(y_i|u_i)\; .
$$ 
Thus $\Wunderlying^n$ corresponds to $n$ independent uses of the
channel~$\Wunderlying$. A key paradigm introduced in \cite{Arikan:09p}
is that of transforming $n$ identical copies (independent uses) of the
channel $\Wunderlying$ into $n$ polar {bit-channels}, through a
successive application of \Arikan\ channel transforms, introduced
shortly. For $i = 0,1,\ldots,n-1$, the $i$-th 
\emph{bit-channel} $\Wbit_i$ has a
binary input alphabet $\calX$, an output alphabet $\calY^n \times \calX^i$,
and transition probabilities defined as follows. 
Let $G$ be the polarization kernel matrix of~\cite{Arikan:09p},
given by 
$$
G 
\ = \
\left[ 
\begin{array}{@{\hspace{0.750ex}}c@{\hspace{1.25ex}}c@{\hspace{0.750ex}}}
1 & 0\\
1 & 1\\ 
\end{array}
\right] \; .
$$
Let $\Gm$ be the $m$-fold Kronecker product of $G$ and let $B_n$ be
the $n \times n$ bit-reversal premutation matrix defined in
\cite[Section\,VII-B]{Arikan:09p}. 
Denote $\uuu_{i-1} = (u_0,u_1,\ldots,u_{i-1})$. 
Then
\begin{multline}
\label{eq:bitChannelDef}
\Wbit_i\bigl(\yyy, \uuu_{i-1}|u_i\bigr)\ \ \deff \\[0.75ex]
\frac{1}{2^{n-1}}\hspace*{-1ex}
\sum_{\vvv \in \mysett{0,1}^{n-1-i}\hspace*{-3ex}} 
\hspace*{-1.00ex}
\Wunderlying^n\Bigl(\yyy\,|\,(\uuu_{i-1},u_i,\vvv) B_n \Gm\Bigr) \; .
\end{multline}

Given the bit-channel output $\yyy$ and $\uuu_{i-1}$, the optimal
(maxim\-um-likelihood) decision rule for estimating $u_i$ is
$$
\widehat{u}_i
\ = \ 
\argmax\bigl\{
\Wbit_i\bigl(\yyy, \uuu_{i-1}|0\bigr),\Wbit_i\bigl(\yyy, \uuu_{i-1}|1\bigr)
\bigr\}
$$
with ties broken arbitrarily. This is the decision rule used
in successive cancellation decoding~\cite{Arikan:09p}. 
As before, we let $\errorProb(\Wbit_i)$ denote the probability that 
$\widehat{u}_i \ne u_i$
under this rule, assuming that the {a priori} distribution
of $u_i$ is $\Bernoulli(1/2)$.

In essence, constructing a polar code of dimension $k$ is equivalent
to finding the $k$ ``best'' bit-channels. In~\cite{Arikan:09p}, 
one is instructed to choose the $k$ bit-channels $\Wbit_i$
with the lowest 
Bhattacharyya bound $Z(\Wbit_i)$ 
on the probability of decision error $\errorProb(\Wbit_i)$.
We note that the
choice of ranking according to these Bhattacharyya bounds stems from the
relative technical ease of manipulating them. A more straightforward
criterion would have been to rank directly according to the probability 
of error $\errorProb(\Wbit_i)$,
and this~is the criterion we will follow here. 

Since $\Wbit_i$ is well defined through
(\ref{eq:bitChannelDef}), this task is indeed explicit, and thus so is
the construction of a polar code. However, note that the output
alphabet size of each bit-channel is exponential in $n$. Thus a
straightforward evaluation of the ranking criterion is intractable for
all but the shortest of codes. Our main objective will be to
circumvent this difficulty.

As a first step towards achieving our goal, we recall that the
bit-channels can be constructed recursively using the \Arikan\ channel
transformations $\Wgeneric\ATbad \Wgeneric$ and $\Wgeneric\ATgood
\Wgeneric$, defined as follows. Let $\Wgeneric\! : \calX \to \calY$ be 
a binary-input, memoryless, symmet\-ric (\BMS) channel.
Then the output alphabet of 
$\Wgeneric\ATbad \Wgeneric$ is $\calY^2$,
the output alphabet of  $\Wgeneric\ATgood \Wgeneric$ 
is $\calY^2 \times \calX$, 
and their transition probabilities are given by
\begin{multline}
\label{eq:badSplit}
\bigl(\Wgeneric {\ATbad} \Wgeneric\bigr)(y_1,y_2 | u_1)\,\ \deff \\
\frac{1}{2} \sum_{u_2 \in \calX} \Wgeneric(y_1|u_1 \xor u_2) \Wgeneric(y_2|u_2)
\end{multline}
and
\begin{multline}
\label{eq:goodSplit}
\bigl(\Wgeneric {\ATgood} \Wgeneric\bigr)(y_1,y_2,u_1|u_2)\,\ \deff \\
\frac{1}{2} \Wgeneric(y_1|u_1 \xor u_2) \Wgeneric(y_2|u_2)
\end{multline}
One consequence of this recursive construction
is that the explosion in the output alphabet size happens 
gradually: each tran\-sform application roughly squares 
the alphabet size. We will take advantage of this fact
in Section~\ref{sec:highLevelAlg}.

\section{Channel Degradation and Upgradation}
\label{sec:degradeUpgrade}
As previously outlined, our solution to the explosion in growth of the output alphabet of $\Wbit_i$ 
is to replace the channel $\Wbit_i$ 
by an approximation. In fact, we will have two approximations, one yielding a ``better'' channel and the other yielding a ``worse'' one. In this section, we formalize these notions.

\begin{figure}
\centering
\subfigure[Degrading]{\label{subfig:degraded}\includegraphics[width=52mm]{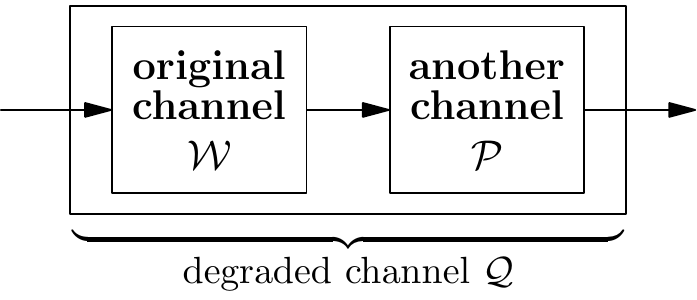}}
\\
\subfigure[Upgrading]{\label{subfig:upgraded}\includegraphics[width=52mm]{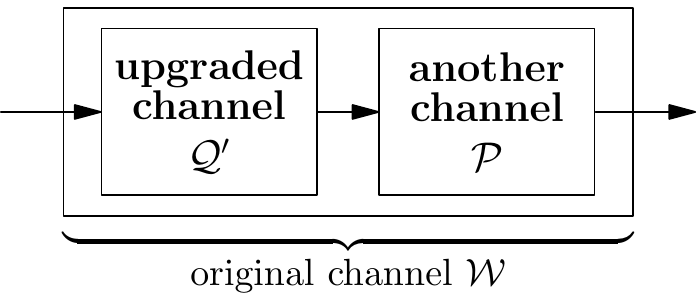}}
\caption{Degrading and upgrading a channel $\Wgeneric$}
\label{fig:degradedUpgraded}
\end{figure}

We say that a channel $\Wdegraded : \calX \to \calYdegraded$ is (stochastically) \emph{degraded} with respect to $\Wgeneric : \calX \to \calY$, if there exists a channel $\intermediateChannel: \calY \to \calYdegraded$ such that for all $\ydegraded \in \calYdegraded$ and $x \in \calX$,
\begin{equation}
\label{eq:degraded}
\Wdegraded(\ydegraded|x) = \sum_{y \in \calY} \Wgeneric(y|x) \cdot \intermediateChannel(\ydegraded|y) \; .
\end{equation}
For a graphical depiction, see Figure~\ref{subfig:degraded}. We write $\Wdegraded \degraded \Wgeneric$ to denote that $\Wdegraded$ is degraded with respect to $\Wgeneric$.

In the interest of brevity and clarity later on, we also define the inverse relation: we say that a channel $\Wupgraded : \calX \to \calYupgraded$ is \emph{upgraded} with respect to $\Wgeneric :\calX \to \calY$ if there exists a channel $\intermediateChannel: \calYupgraded \to \calY$ such that for all $\yupgraded \in \calYupgraded$ and $x \in \calX$,
\begin{equation}
\label{eq:upgraded}
\Wgeneric(y|x) = \sum_{\yupgraded \in \calYupgraded} \Wupgraded(\yupgraded|x) \cdot \intermediateChannel(y|\yupgraded) 
\end{equation}
(see Figure~\ref{subfig:upgraded}). Namely, $\Wupgraded$ can be degraded to $\Wgeneric$. Similarly, we write this as $\Wupgraded \upgraded \Wgeneric$.

By definition,
\begin{equation}
\label{eq:degradedUpgradedIff}
\Wgeneric \degraded \Wgeneric' \quad \mbox{if and only if} \quad \Wgeneric' \upgraded \Wgeneric \; .
\end{equation}
Also, it is easily shown that ``degraded'' is a transitive relation:
\begin{equation}
\label{eq:degradedTransitive}
\mbox{If} \quad \Wgeneric \degraded \Wgeneric' \quad \mbox{and} \quad \Wgeneric' \degraded \Wgeneric'' \quad \mbox{then} \quad \Wgeneric \degraded \Wgeneric'' \; .
\end{equation}
Thus, the ``upgraded'' relation is transitive as well.
Lastly, since a channel is both degraded and upgraded with respect to itself (take the intermediate channel as the identity function), we have that both relations are reflexive:
\begin{equation}
\label{eq:upgradedDegradedReflexiv}
\Wgeneric \degraded \Wgeneric \quad \mbox{and} \quad \Wgeneric \upgraded \Wgeneric \; . 
\end{equation}

If a channel $\Wequivalent$ is both degraded and upgraded with respect to $\Wgeneric$, then we say that $\Wgeneric$ and $\Wequivalent$ are \emph{equivalent}, and denote this by $\Wgeneric \equivalent \Wequivalent$. Since ``degraded'' and ``upgraded'' are transitive relations, it follows 
that the ``equivalent'' relation is transitive as well.
Also, by (\ref{eq:degradedUpgradedIff}), we have that ``equivalent'' is a symmetric relation:
\begin{equation}
\label{eq:equivalentSymmetric}
\Wgeneric \equivalent \Wgeneric'  
\quad \mbox{if and only if} \quad 
\Wgeneric' \equivalent \Wgeneric \; .
\end{equation}
Lastly, since a channel $\Wgeneric$ is both upgraded and degraded with respect to itself, we have by (\ref{eq:upgradedDegradedReflexiv}) that ``equivalent'' is a reflexive relation.
Thus, channel equivalence is indeed an equivalence relation.

Let $\Wgeneric:\calX \to \calY$ be a given BMS channel. We now set the notation for three quantities of interest.
i) Denote by $\errorProb(\Wgeneric)$ the probability of error under maximum-likelihood decision, where ties are broken arbitrarily, and the input distribution is $\Bernoulli(1/2)$. That is,
\begin{equation}
\label{eq:errorProbDef}
\errorProb(\Wgeneric) = 
\frac{1}{2} \sum_{y \in \calY} \min\{\Wgeneric(y|0),\Wgeneric(y|1)\} \; .
\end{equation}
ii) Denote by $Z(\Wgeneric)$ the Bhattacharyya parameter,
\begin{equation}
\label{eq:BhattacharyyaDef}
Z(\Wgeneric) = \sum_{y \in \calY} \sqrt{\Wgeneric(y|0)\Wgeneric(y|1)} \; .
\end{equation}
iii) Denote by $I(\Wgeneric)$ the capacity,
\[
I(\Wgeneric) = \sum_{y \in \calY} \sum_{x \in \calX} \frac{1}{2} \Wgeneric(y|x) \log \frac{\Wgeneric(y|x)}{\frac{1}{2}\Wgeneric(y|0)+\frac{1}{2}\Wgeneric(y|1)} \; .
\]
The following lemma states that these three quantities behave as expected with respect to the degrading and upgrading relations. The equation most important to us will be (\ref{eq:errorProbDegrading}).

\begin{lemm}[{\cite[page 207]{RichardsonUrbanke:08b}}]
\label{lemm:degradingAndEZI}
Let $\Wgeneric:\calX \to \calY$ be a \BMS\ channel and let $\Wdegraded : \calX \to \calYdegraded$ be degraded with respect to $\Wgeneric$, that is, $\Wdegraded \degraded \Wgeneric$. Then,
\begin{align}
\label{eq:errorProbDegrading}
\errorProb(\Wdegraded) & \geq \errorProb(\Wgeneric) \; , \\
\label{eq:BhattacharyyaProbDegrading}
\quad Z(\Wdegraded) & \geq Z(\Wgeneric) \; , \quad \mbox{and} \\
\label{eq:capacityDegrading}
\quad I(\Wdegraded) & \leq  I(\Wgeneric) \; .
\end{align}

Moreover, all of the above continues to hold if we replace ``degraded'' by ``upgraded'', $\degraded$ by $\upgraded$, and reverse the inequalities. Specifically, if $\Wgeneric \equiv \Wdegraded$, then the weak inequalities are in fact equalities.
\end{lemm}

\begin{proof}
We consider only the first part, since the ``Moreover'' part follows easily from Equation~(\ref{eq:degradedUpgradedIff}). 
For a simple proof of (\ref{eq:errorProbDegrading}), recall the definition of degradation (\ref{eq:degraded}), and note that 
\begin{multline*}
\errorProb(\Wdegraded) = \frac{1}{2}\sum_{z \in \calZ} \min\{\Wdegraded(\calZ|0),\Wdegraded(\calZ|1)\} = \\
\frac{1}{2}\sum_{z \in \calZ} \min\left\{\sum_{y \in \calY} \Wgeneric(y|0) \cdot \intermediateChannel(\ydegraded|y),\sum_{y \in \calY} \Wgeneric(y|1) \cdot \intermediateChannel(\ydegraded|y)\right\} \\
\geq \frac{1}{2}\sum_{z \in \calZ} \sum_{y \in \calY} \min\{\Wgeneric(y|0),\Wgeneric(y|1)\} \cdot \intermediateChannel(\ydegraded|y) = \errorProb(\Wgeneric)
\end{multline*}
Equation (\ref{eq:BhattacharyyaProbDegrading}) is concisely proved in \cite[Lemma 1.8]{Korada:09z}. Equation (\ref{eq:capacityDegrading}) is a simple consequence of the data-processing inequality \cite[Theorem 2.8.1]{CoverThomas:06b}.
\end{proof}

Note that it may be the case that $y$ is its own conjugate. That is, $y$ and $\bary$ are the same symbol (an erasure). It would make our proofs simpler if this special case was assumed not to happen. We will indeed assume this later on, with the next lemma providing most of the justification.

\begin{lemm}
\label{lemm:allPairs}
Let $\Wgeneric : \calX \to \calY$ be a \BMS\ channel. There exists a \BMS\ channel $\Wequivalent : \calX \to \calZ$
such that i) $\Wequivalent$ is equivalent to $\Wgeneric$, and ii) for all $z \in \calZ$ we have that $z$ and $\bar{z}$ are distinct.
\end{lemm}

\begin{proof}
If $\Wgeneric$ is such that for all $y \in \calY$ we have that $y$ and $\bary$ are distinct, then we are done, since we can take $\Wequivalent$ equal to $\Wgeneric$.

Otherwise, let $\yerasure \in \calY$ be such that $\yerasure$ and $\baryerasure$ are the same symbol. Let the alphabet $\calZ$ be defined as follows:
\[
\calZ = \left(\calY \setminus \mysett{\yerasure}\right) \cup \mysett{z_1,z_2} \; ,
\]
where $z_1$ and $z_2$ are new symbols, not already in $\calY$. Now, define the channel $\Wequivalent:\calX \to \calZ$ as follows. For all $z \in \calZ$ and $x \in \calX$,
\[
\Wequivalent(z|x) =
\begin{cases}
\Wgeneric(z|x) & \mbox{if $z \in \calY$,} \\
\frac{1}{2} \Wgeneric(\yerasure|x) & \mbox{if $z = z_1$ or $z = z_2$.}
\end{cases}
\]

We first show that $\Wequivalent \upgraded \Wgeneric$. To see this, take the intermediate channel $\intermediateChannel : \calZ \to \calY$ as the channel that maps (with probability 1) $z_1$ and $z_2$ to $\yerasure$, and all other symbols to themselves. Next, we show that $\Wequivalent \degraded \Wgeneric$. To see this, define the intermediate channel $\intermediateChannel : \calY \to \calZ$ as follows.
\[
\intermediateChannel(z|y) =
\begin{cases}
1 & \mbox{if $z=y$,} \\
\frac{1}{2} & \mbox{if $y = \yerasure$ and $z \in \mysett{z_1,z_2}$,} \\
0 & \mbox{otherwise.}
\end{cases}
\]

To sum up, we have constructed a new channel $\Wequivalent$ which is equivalent to $\Wgeneric$, and contains one less self-conjugate symbol ($\yerasure$ was replaced by the pair $z_1,z_2$). It is also easy to see that $\Wequivalent$ is \BMS. We can now apply this construction over and over, until the resulting channel has no self-conjugate symbols.
\end{proof}

Now that Lemma~\ref{lemm:allPairs} is proven, we will indeed assume from this point forward that all channels are \BMS\ and have no output symbols $y$ such that $y$ and $\bary$ are equal. As we will show later on, this assumption does not limit us. Moreover, given a generic \BMS\ channel $\Wgeneric: \calX \to \calY$, we will further assume that for all $y \in \calY$, at least one of the probabilities $\Wgeneric(y|0)$ and $\Wgeneric(\bary|0)$ is positive (otherwise, we can remove the pair of symbols $y,\bary$ from the alphabet, since they can never occur).

Given a channel $\Wgeneric : \calX \to \calY$, we now define for each output symbol $y \in \calY$ an associated \emph{likelihood ratio}, denoted $\LR_\Wgeneric(y)$. Specifically,
\[
\LR_\Wgeneric(y) = \frac{\Wgeneric(y|0)}{\Wgeneric(y|1)} =
\frac{\Wgeneric(y|0)}{\Wgeneric(\bar{y}|0)}
\]
(if $\Wgeneric(\bar{y}|0) = 0$, then we must have by assumption that $\Wgeneric(y|0) > 0$, and we define $\LR_\Wgeneric(y) = \infty$). If the channel $\Wgeneric$ is understood from the context, we will abbreviate $\LR_\Wgeneric(y)$ to $\LR(y)$.

\section{High-Level Description of the Algorithms}
\label{sec:highLevelAlg}

\looseness=-1
In this section, we give a high level description of our algorithms for approximating a bit channel. We then show how these approximations can be used in order to construct a polar code.

In order to completely specify the approximating algorithms, one has to supply two \emph{merging functions}, a degrading merging function \degradingMerge and an upgraded merging function \upgradingMerge. We will now define the properties required of our merging functions, leaving the specification of the functions we have actually used to the next section. The next section will also make clear why we have chosen to call these functions ``merging''.

For a degrading merge function \degradingMerge, the following must hold. For a \BMS\ channel $\Wgeneric$ and positive integer $\mu$, the output of $\degradingMerge(\Wgeneric,\mu)$ is a \BMS\ channel $\Wdegraded$ such that i) $\Wdegraded \degraded \Wgeneric$ is degraded with respect to $\Wgeneric$, and ii) The size of the output alphabet of $\Wdegraded$ is at most $\mu$. We define the properties required of $\upgradingMerge$ similarly, but with ``degraded'' replaced by ``upgraded'' and $\degraded$ by $\upgraded$.

Let $0 \leq i < n$ be an integer with binary representation $i=\binaryRep{b_1,b_2,\ldots,b_m}$, where $b_1$ is the most significant bit. Algorithms~\ref{alg:highLevelDegrade} and \ref{alg:highLevelUpgrade} contain our procedures for finding a degraded and upgraded approximation of the bit channel $\Wbit_i^{(m)}$, respectively. In words, we employ the recursive constructions (\ref{eq:badSplit}) and (\ref{eq:goodSplit}), taking care to reduce the output alphabet size of each intermediate channel from at most $2 \mu^2$ (apart possibly from the underlying channel $\Wunderlying$) to at most $\mu$. 

\begin{algorithm}
\SetInd{0.49em}{0.49em}
\caption{Bit-channel degrading procedure}
\label{alg:highLevelDegrade}
\Input{An underlying \BMS\ channel $\Wunderlying$, a bound $\mu=2\outputPairsCount$ on the output alphabet size, a code length $n=2^m$, an index $i$ with binary representation $i=\binaryRep{b_1,b_2,\ldots,b_m}$.}
\Output{A \BMS\ channel that is degraded with respect to the bit channel $\Wbit_i$.}
\label{algline:QinitDegrade} $\QAlg \leftarrow \degradingMerge(\Wunderlying,\mu)$\;
\For{$j = 1,2,\ldots, m$}
{
  \eIf{$b_j=0$}
  {
    $\WAlg \leftarrow \QAlg \ATbad \QAlg$
  }
  {
    $\WAlg \leftarrow \QAlg \ATgood \QAlg$
  }
  $\QAlg \leftarrow \degradingMerge(\WAlg,\mu)$\;
}
\Return{$\QAlg$}\;
\end{algorithm}

\begin{algorithm}
\SetInd{0.49em}{0.49em}
\caption{Bit-channel upgrading procedure}
\label{alg:highLevelUpgrade}
\Input{An underlying \BMS\ channel $\Wunderlying$, a bound $\mu=2\outputPairsCount$ on the output alphabet size, a code length $n=2^m$, an index $i$ with binary representation $i=\binaryRep{b_1,b_2,\ldots,b_m}$.}
\Output{A \BMS\ channel that is upgraded with respect to the bit channel $\Wbit_i$.}
$\QprimeAlg \leftarrow \upgradingMerge(\Wunderlying,\mu)$\;
\For{$j = 1,2,\ldots, m$}
{
  \eIf{$b_j=0$}
  {
    $\WAlg \leftarrow \QprimeAlg \ATbad \QprimeAlg$
  }
  {
    $\WAlg \leftarrow \QprimeAlg \ATgood \QprimeAlg$
  }
  $\QprimeAlg \leftarrow \upgradingMerge(\WAlg,\mu)$\;
}
\Return{$\QprimeAlg$}\;
\end{algorithm}

The key to proving the correctness of Algorithms~\ref{alg:highLevelDegrade} and \ref{alg:highLevelUpgrade} is the following lemma. It is essentially a restatement of \cite[Lemma 4.7]{Korada:09z}. For completeness, we restate the proof as well.

\begin{lemm}
\label{lemm:degradedPreserved}
Fix a binary input channel $\Wgeneric:\calX \to \calY$, and denote 
\[
\Wbad = \Wgeneric \ATbad \Wgeneric \; , \quad  \Wgood = \Wgeneric \ATgood \Wgeneric \; .
\]
Next, let $\Wdegraded \degraded \Wgeneric$ be a degraded with respect to $\Wgeneric$, and denote
\[
\Qbad = \Wdegraded \ATbad \Wdegraded \; , \quad  \Qgood = \Wdegraded \ATgood \Wdegraded \; .
\]
Then,
\[
\Qbad \degraded \Wbad \quad \mbox{and} \quad \Qgood \degraded \Wgood \; .
\]
Namely, the degradation relation is preserved by the channel transformation operation. 

Moreover, all of the above continues to hold if we replace ``degraded'' by ``upgraded'' and $\degraded$ by $\upgraded$.
\end{lemm}
\begin{proof}
We will prove only the ``degraded'' part, since it implies the ``upgraded'' part (by interchanging the roles of $\Wgeneric$ and $\Wdegraded$).

Let $\intermediateChannel : \calY \to \calZ$ be the channel which degrades $\Wgeneric$ to $\Wdegraded$: for all $z \in \calZ$ and $x \in \calX$,
\begin{equation}
\label{eq:degradedWtoTildeW}
\Wdegraded(z|x) = \sum_{y \in \calY} \Wgeneric(y|x)\intermediateChannel(z|y) \; .
\end{equation}

We first prove $\Qbad \degraded \Wbad$. By (\ref{eq:badSplit}) applied to $\Wdegraded$, we get that for all $(z_1,z_2) \in \calZ^2$ and $u_1 \in \calX$,
\[
\Qbad((z_1,z_2)|u_1) = \sum_{u_2 \in \calX} \frac{1}{2} \Wdegraded(z_1|u_1 \xor u_2) \Wdegraded(z_2|u_2) \; .
\]
Next, we expand $\Wdegraded$ twice according to (\ref{eq:degradedWtoTildeW}), and get
\begin{multline*}
\Qbad((z_1,z_2)|u_1) = \\
\sum_{(y_1,y_2) \in \calY^2} \sum_{u_2} 
 \frac{1}{2} W(y_1|u_1 \xor u_2) W(y_2 | u_2) \intermediateChannel(z_1|y_1) \intermediateChannel(z_2|y_2) \; .
\end{multline*}
By (\ref{eq:badSplit}), this reduces to
\begin{multline}
\label{eq:WprimeHalfWay}
\Qbad((z_1,z_2)|u_1) = \\
\sum_{(y_1,y_2) \in \calY^2} \Wbad((y_1,y_2)|u_1) \intermediateChannel(z_1|y_1) \intermediateChannel(z_2|y_2) \; .
\end{multline}
Next, define the channel $\intermediateChannel^*:\calY^2 \to \calZ^2$ as follows. For all $(y_1,y_2) \in \calY^2$ and $(z_1,z_2) \in \calZ^2$,
\[
\intermediateChannel^*((z_1,z_2)|(y_1,y_2)) = \intermediateChannel(z_1|y_1) \intermediateChannel(z_2|y_2) \; .
\]
It is easy to prove that $\intermediateChannel^*$ is indeed a channel (we get a probability distribution on $\calZ^2$ for every fixed $(y_1,y_2) \in \calY^2$). Thus, (\ref{eq:WprimeHalfWay}) reduces to
\begin{multline*}
\Qbad((z_1,z_2)|u_1) = \\
\sum_{(y_1,y_2) \in \calY^2} \Wbad((y_1,y_2)|u_1) \intermediateChannel^*((z_1,z_2)|(y_1,y_2)) \; ,
\end{multline*}
and we get by (\ref{eq:degraded}) that $\Qbad \degraded \Wbad$. The claim $\Qgood \degraded \Wgood$ is proved in much the same way.
\end{proof}

\begin{prop}
\label{prop:algsDegradingUpgrading}
The output of Algorithm~\ref{alg:highLevelDegrade} (Algorithm~\ref{alg:highLevelUpgrade}) is a \BMS\ channel that is degraded (upgraded) with respect to $\Wbit_i^{(m)}$.
\end{prop}

\begin{proof}
The proof follows easily from Lemma~\ref{lemm:degradedPreserved}, by induction on $j$.
\end{proof}

Recall that, ideally, a polar code is constructed as follows. We are given an underlying channel $\Wunderlying :\calX \to \calY$, a specified codeword length $n = 2^m$, and a target block error rate $\errorBlock$. We choose the largest possible subset of bit-channels $\Wbit_i$ such that the sum of their probabilities of error $\errorProb(\Wbit_i)$ is not greater than $\errorBlock$. The resulting code is spanned by the rows in $B_n \Gm$ corresponding to the subset of chosen bit-channels. Denote the rate of this code as $\Rexact$.

Since we have no computational handle on the bit channels $\Wbit_i$, we must resort to approximations. Let $\Wdegraded_i$ be the result of running Algorithm~\ref{alg:highLevelDegrade} on $\Wunderlying$ and $i$. Since $\Wdegraded_i \degraded \Wbit_i$, we have by (\ref{eq:errorProbDegrading}) that $\errorProb(\Wdegraded_i) \geq \errorProb(\Wbit_i)$. Note that since the output alphabet of $\Wdegraded_i$ is small (at most $\mu$), we can actually compute $\errorProb(\Wdegraded_i)$. We now mimic the ideal construction by choosing the largest possible subset of indices for which the sum of $\errorProb(\Wdegraded_i)$ is at most $\errorBlock$. Note that for this subset we have that the sum of $\errorProb(\Wbit_i)$ is at most $\errorBlock$ as well. Thus, the code spanned by the corresponding rows of $B_n \Gm$ is assured to have block error probability of at most $\errorBlock$.

Denote the rate of this code by $\Rdegraded$. It is easy to see that $\Rdegraded \leq \Rexact$. In order to gauge the difference between the two rates, we compute a third rate, $\Rupgraded$, such that  $\Rupgraded \geq \Rexact$ and consider the difference $\Rupgraded - \Rdegraded$. The rate $\Rupgraded$ is computed the same way that $\Rdegraded$ is, but instead of using Algorithm~\ref{alg:highLevelDegrade} we use Algorithm~\ref{alg:highLevelUpgrade}. Recall from Figure~\ref{fig:BSCplots} that $\Rdegraded$ and $\Rupgraded$ are typically very close.

We end this section by noting a point that will be needed in the proof of Theorem~\ref{theo:totalRunningTime} below. Consider the running time needed in order to approximate all $n$ bit channels. Assume that each invocation of either $\degradingMerge$ or $\upgradingMerge$ takes time $\tau=\tau(\mu)$. Thus, the time needed for approximating a single bit channel using either Algorithm~\ref{alg:highLevelDegrade} or Algorithm~\ref{alg:highLevelUpgrade} is $O(m \tau)$. A naive analysis suggests that the time needed in order to approximate all $n$ bit-channels is $O(n \cdot m \tau)$. However, significant savings can be gained by noticing that intermediate calculations can be shared between bit-channels. For example, in a naive implementation we would approximate $\Wunderlying \ATbad \Wunderlying$ over and over again, $n/2$ times instead of only once. A quick calculation shows that the number of \emph{distinct} channels one needs to approximate is $2n-1-1$. That is, following both branches of the ``if'' statement of (without loss of generality) Algorithm~\ref{alg:highLevelDegrade} would produce $2^j$ channels for each level $1 \leq j \leq m$. Thus, the total running time can be reduced to $O((2n-2)\cdot \tau)$, which is simply $O(n \cdot \tau)$.

\section{Merging Functions}
\label{sec:mergingFunctions}
In this section, we specify the degrading and upgrading functions used to reduce the output alphabet size. These functions are referred to as $\degradingMerge$ and $\upgradingMerge$ in Algorithms \ref{alg:highLevelDegrade} and \ref{alg:highLevelUpgrade}, respectively. For now, let us treat our functions as heuristic (delaying their analysis to Section~\ref{sec:analysis}).

\begin{figure*}
\hfill
\subfigure[]{\label{subfig:degradingWQ}\includegraphics[width=72mm]{{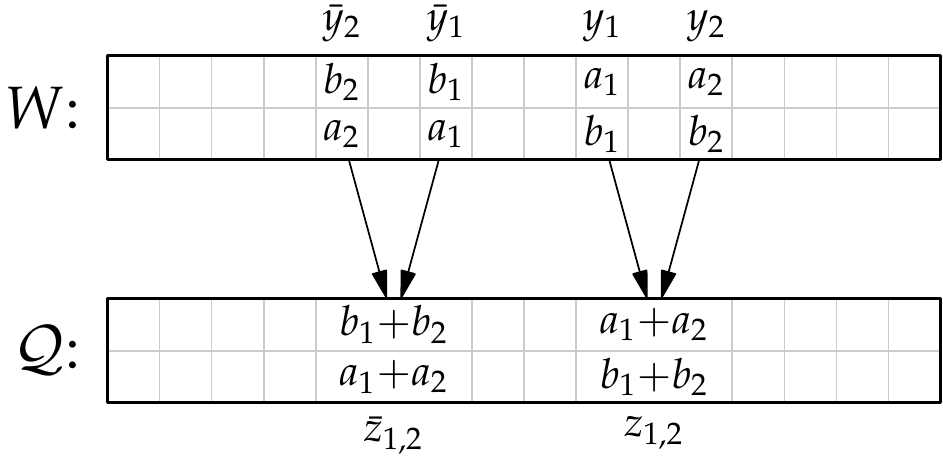}}}
\hfill\hfill
\subfigure[]{\label{subfig:degradingP}\includegraphics[width=72mm]{{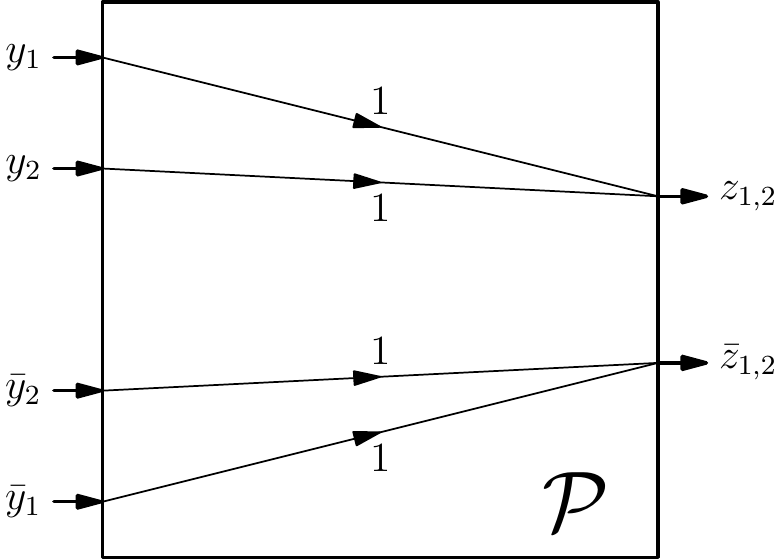}}}
\hfill{}
\caption{Degrading $\Wgeneric$ to $\Wdegraded$. (a) The degrading merge operation: the entry in the first/second row of a channel is the probability of receiving the corresponding symbol, given that a $0$/$1$ was transmitted. (b) The intermediate channel $\intermediateChannel$.}
\label{fig:degradingWQP}
\end{figure*}

\subsection{Degrading-merge function}
We first note that the problem of degrading a binary-input channel to a channel with a prescribed output alphabet size was independently considered by Kurkoski and Yagi \cite{KurkoskiYagi:11a}. The main result in \cite{KurkoskiYagi:11a} is an optimal degrading strategy, in the sense that the capacity of the resulting channel is the largest possible. In this respect, the method we now introduce is sub-optimal. However, as we will show, the complexity of our method is superior to that presented in \cite{KurkoskiYagi:11a}.

The next lemma shows how one can reduce the output alphabet size by 2, and get a degraded channel. It is our first step towards defining a valid $\degradingMerge$ function.
\begin{lemm}
\label{lemm:degradeStep}
Let $\Wgeneric:\calX \to \calY$ be a BMS channel, and let $y_1$ and $y_2$ be symbols in the output alphabet $\calY$. Define the channel $\Wdegraded : \calX \to \calZ$ as follows (see Figure~\ref{subfig:degradingWQ}). The output alphabet $\calZ$ is given by
\[
\calZ = \calY \setminus \mysett{y_1,\bary_1,y_2,\bary_2} \cup \mysett{\zmerged,\barzmerged}\; .
\]
For all $x \in \calX$ and $z \in \calZ$, define
\[
\Wdegraded(z|x) =
\begin{cases}
\Wgeneric(z|x) & \mbox{if $z \not\in \mysett{\zmerged,\barzmerged}$,} \\
\Wgeneric(y_1|x) + \Wgeneric(y_2|x) & \mbox{if $z = \zmerged$,} \\
\Wgeneric(\bary_1|x) + \Wgeneric(\bary_2|x) & \mbox{if $z = \barzmerged$.}
\end{cases}
\]
Then $\Wdegraded \degraded \Wgeneric$. That is, $\Wdegraded$ is degraded with respect to $\Wgeneric$.
\end{lemm}
\begin{proof}
Take the intermediate channel $\intermediateChannel : \calY \to \calZ$ as the channel that maps with probability 1 as follows (see Figure~\ref{subfig:degradingP}): both $y_1$ and $y_2$ map to $\zmerged$, both $\bary_1$ and $\bary_2$ map to $\barzmerged$, other symbols map to themselves. Recall that we have assumed that $\Wgeneric$ does not contain an erasure symbol, and this continues to hold for $\Wdegraded$.
\end{proof}

We now define the $\degradingMerge$ function we have used. It gives good results in practice and is amenable to a fast implementation. Assume we are given a BMS channel $\Wgeneric: \calX \to \calY$ with an alphabet size of $2L$ (recall our assumption of no self-conjugates), and wish to
transform $\Wgeneric$ into a degraded version of itself while 
reducing its alphabet size to $\mu$. If $2L \leq \mu$, then we are done, since we can take the degraded version of $\Wgeneric$ to be $\Wgeneric$ itself. Otherwise, we do the following. Recall that for each $y$ we have that $\LR(y) = 1/\LR(\bary)$, where in this context $1/0=\infty$ and $1/\infty = 0$. Thus, our first step is to choose from each pair $(y,\bary)$ a representative such that $\LR(y) \geq 1$. Next, we order these $L$ representative such that
\begin{equation}
\label{eq:yOrdered}
1 \leq \LR(y_1) \leq \LR(y_2) \leq \cdots \leq \LR(y_L) \; .
\end{equation}
We now ask the following
: for which index $1 \leq i \leq L-1$ does the channel resulting from the application of Lemma~\ref{lemm:degradeStep} to $\Wgeneric$, $y_i$, and $y_{i+1}$ result in a channel with largest capacity? Note that instead of considering $\binom{L}{2}$ merges, we consider only $L-1$. After finding the maximizing index $i$ we indeed apply Lemma~\ref{lemm:degradeStep} and get a degraded channel $\Wdegraded$ with an alphabet size smaller by $2$ than that of $\Wgeneric$. The same process is applied to $\Wdegraded$, until the output alphabet size is not more than $\mu$.

\begin{algorithm}
\SetInd{0.49em}{0.49em}
\caption{The \protect\degradingMerge function}
\label{alg:degradeChannel}
\Input{A \BMS\ channel $\Wgeneric:\calX \to \calY$ where $\sizee{\calY} = 2\inputPairsCount$, a bound $\mu=2\outputPairsCount$ on the output alphabet size.}
\Output{A degraded channel $\Wdegraded : \calX \to \calY'$, where $\sizee{\calY'} \leq \mu$.}
\tcp{Assume $1 \leq \LR(y_1) \leq \LR(y_2) \leq \cdots \leq \LR(y_L)$}
\For{$i = 1,2,\ldots, L-1$}
{
$\datum \leftarrow$ new data element\;
$\datum.a \leftarrow \Wgeneric(y_i|0) \; , \quad \datum.b \leftarrow \Wgeneric(\bary_i|0)$\;
$\datum.a' \leftarrow \Wgeneric(y_{i+1}|0) \; , \quad \datum.b' \leftarrow \Wgeneric(\bary_{i+1}|0)$\;
$\datum.\deltaI \leftarrow \calcDeltaI(\datum.a,\datum.b,\datum.a',\datum.b')$\;
$\insertRightmost(\datum)$\; 
}
$\ell = L$\;
\While{$\ell > \outputPairsCount$}
{
  $\datum \leftarrow \getMin()$\;
  $a^+ = \datum.a+\datum.a' \; , \quad b^+ = \datum.b+\datum.b'$\;
  $\datumLeft \leftarrow \datum.\leftAlg$\;
  $\datumRight \leftarrow \datum.\rightAlg$\;
  $\removeMin()$\;
  $\ell \leftarrow \ell -1$\;
  \If{$\datumLeft \neq \nullAlg$}
  {
    $\datumLeft.a' = a^+$\;
    $\datumLeft.b' = b^+$\;
    $\datumLeft.\deltaI \leftarrow \calcDeltaI(\datumLeft.a,\datumLeft.b,a^+,b^+)$\;
    $\valueUpdated(\datumLeft)$\;
  }
  \If{$\datumRight \neq \nullAlg$}
  {
    $\datumRight.a = a^+$\;
    $\datumRight.b = b^+$\;
    $\datumRight.\deltaI \leftarrow \calcDeltaI(a^+,b^+,\datumRight.a',\datumRight.b')$\;
    $\valueUpdated(\datumRight)$\;
  }
}
Construct $\Wdegraded$ according to the probabilities in the data structure and return it.
\end{algorithm}

In light of Lemma~\ref{lemm:degradeStep} and (\ref{eq:yOrdered}), a simple yet important point to note is that if $y_i$ and $y_{i+1}$ are merged to $z$, then
\begin{equation}
\label{eq:LRMean}
\LR(y_i) \leq \LR(z) \leq \LR(y_{i+1}) \; .
\end{equation}
Namely, the original LR ordering is essentially preserved by the merging operation. Algorithm~\ref{alg:degradeChannel} contains an implementation of our merging procedure. It relies on the above observation in order to improve complexity and runs in $O(L \cdot \log L)$ time. Thus, assuming $L$ is at most $2\mu^2$, the running time of our algorithm is $O(\mu^2 \log \mu)$. In contrast, had we used the degrading method presented in  \cite{KurkoskiYagi:11a}, the running time would have been $O(\mu^5)$.

\begin{figure}[t]
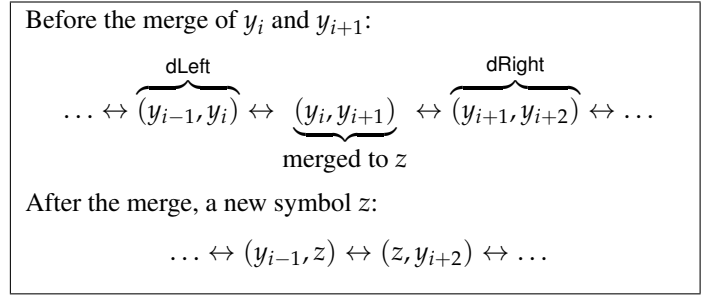

\fbox{
\parbox{\linewidth}{
Before the merge of $y_i$ and $y_{i+1}$:
\[
\ldots \leftrightarrow \overbrace{(y_{i-1},y_i)}^{\datumLeft} \leftrightarrow \underbrace{(y_i,y_{i+1})}_{\mbox{merged to $z$}} \leftrightarrow \overbrace{(y_{i+1},y_{i+2})}^{\datumRight} \leftrightarrow \ldots
\]
After the merge, a new symbol $z$:
\[
\ldots \leftrightarrow (y_{i-1},z) \leftrightarrow (z,y_{i+2}) \leftrightarrow \ldots
\]
}}
\caption{Graphical depiction of the doubly-linked-list before and after a merge.}
\label{fig:mergeBeforeAfter}
\end{figure}

Our implementation assumes an underlying data structure and data elements as follows. Our data structure stores data elements, where each data element corresponds to a pair of adjacent letters $y_i$ and $y_{i+1}$, in the sense of the ordering in (\ref{eq:yOrdered}). Each data element has the following fields:
\[
a\;,\quad
b\;,\quad
a'\;,\quad
b'\;,\quad
\deltaI\;,\quad
\datumLeft\;,\quad
\datumRight\;,\quad
h \; .
\]
The fields $a$, $b$, $a'$, and $b'$ store the probabilities $\Wgeneric(y_i|0)$, $\Wgeneric(\bary_i|0)$, 
$\Wgeneric(y_{i+1}|0)$, and $\Wgeneric(\bary_{i+1}|0)$, respectively. The field $\deltaI$ contains the difference in capacity that would result from applying Lemma~\ref{lemm:degradeStep} to $y_{i}$ and $y_{i+1}$. Note that $\deltaI$ is only a function of the above four probabilities, and thus the function $\calcDeltaI$ used to initialize this field is given by
\[
\calcDeltaI(a,b,a',b')=C(a,b)+C(a',b')-C(a^+,b^+) \; ,
\]
where
\[
C(a,b) = -(a+b) \log_2((a+b)/2)+a \log_2(a) +b \log_2(b) \; ,
\]
we use the shorthand
\[
a^+ = a + a' \; , \quad b^+ = b + b' \; ,
\]
and $0 \log_2 0$ is defined as $0$. The field $\datumLeft$ is a pointer to the data element corresponding to the pair $y_{i-1}$ and $y_i$ (or ``null'', if $i=1$). Likewise, $\datumRight$ is a pointer to the element corresponding to the pair $y_{i+1}$ and $y_{i+2}$ (see Figure~\ref{fig:mergeBeforeAfter} for a graphical depiction). Apart from these, each data element contains an integer field $h$, which will be discussed shortly.


We now discuss the functions that are the interface to our data structure: $\insertRightmost$, $\getMin$, $\removeMin$, and $\valueUpdated$. Our data structure combines the attributes of a doubly-linked-list \cite[Section 10.2]{CLRS:01b} and a heap\footnote{In short, a heap is a data structure that supports four operations: ``insert'', ``getMin'', ``removeMin'', and ``valueUpdated''. In our implementation, the running time of ``getMin'' is constant, while the running time of the other operations is logarithmic in the heap size.} \cite[Chapter 6]{CLRS:01b}. The doubly-linked list is implemented through the $\datumLeft$ and $\datumRight$ fields of each data element, as well as a pointer to the rightmost element of the list. 
Our heap will have the ``array'' implementation, as described in \cite[Section 6.1]{CLRS:01b}. Thus, each data element will have a corresponding index in the heap array, and this index is stored in the field $h$. The doubly-linked-list will be ordered according to the corresponding $\LR$ value, while the heap will be sorted according to the $\deltaI$ field. 

\label{subsec:mergingFunctions:degrade}

The function $\insertRightmost$ inserts a data element as the rightmost element of the list and updates the heap accordingly. The function $\getMin$ returns the data element with smallest $\deltaI$. Namely, the data element corresponding to the pair of symbols we are about to merge.
The function $\removeMin$ removes the element returned by $\getMin$ from both the linked-list and the heap. The function $\valueUpdated$ updates the heap due to a change in $\deltaI$ resulting from a merge, but does not change the linked list in view of (\ref{eq:LRMean}).

The running time of $\getMin$ is $O(1)$, and this is obviously also the case for $\calcDeltaI$.  Due to the need of updating the heap, the running time of $\removeMin$, $\valueUpdated$, and $\insertRightmost$ is $O(\log L)$. The time needed for the initial sort of the $\LR$ pairs is $O(L \cdot \log L)$. Hence, since the initializing for-loop in Algorithm~\ref{alg:degradeChannel} has $L$ iterations and the while-loop has $L-\outputPairsCount$ iterations, the total running time of Algorithm~\ref{alg:degradeChannel} is $O(L \cdot \log L)$.

Note that at first sight, it may seem as though there might be an even better heuristic to employ. As before, assume that the $y_i$ are ordered according to their likelihood ratios, and all of these are at least $1$. Instead of limiting the application of Lemma~\ref{lemm:degradeStep} to $y_i$ and $y_{i+1}$, we can broaden our search and consider the penalty in capacity incurred by merging arbitrary $y_i$ and $y_j$, where $i \neq j$. Indeed, we could further consider merging arbitrary $y_i$ and $\bary_j$, where $i \neq j$. Clearly, this broader search will result in worse complexity. However, as the next theorem shows, we will essentially gain nothing by it.

\begin{theo}
\label{theo:degradeOptimal}
Let $\Wgeneric:\calX \to \calY$ be a BMS channel, with 
\[
\calY = \mysett{y_1,y_2,\ldots,y_L,\bary_1,\bary_2,\ldots,\bary_L} \; .
\]
Assume that
\[
1 \leq \LR(y_1) \leq \LR(y_2) \leq \cdots \leq \LR(y_L) \; .
\]
For symbols $\w_1,\w_2 \in \calY$, denote by $I(\w_1,\w_2)$ the capacity of the channel one gets by the application of Lemma~\ref{lemm:degradeStep} to $\w_1$ and $\w_2$. Then, for all distinct $1 \leq i \leq L$ and $1 \leq j \leq L$,
\begin{equation}
\label{eq:theo:degradeOptimal_LRgeq1}
I(\bary_i,\bary_j) = I(y_i,y_j) \geq I(y_i,\bary_j) = I(\bary_i,y_j)\; .
\end{equation}
Moreover, for all $1 \leq i < j < k \leq L$ we have that either
\[
I(y_i,y_j) \geq I(y_i,y_k)\; ,
\]
or
\[
I(y_j,y_k) \geq I(y_i,y_k)\; .
\]
\end{theo}

We note that Theorem~\ref{theo:degradeOptimal} seems very much related to \cite[Lemma 5]{KurkoskiYagi:11a}. However, one important difference is that Theorem~\ref{theo:degradeOptimal} deals with the case in which the degraded channel is constrained to be symmetric, while \cite[Lemma 5]{KurkoskiYagi:11a} does not. At any rate, for completeness, we will prove Theorem~\ref{theo:degradeOptimal} in Appendix~\ref{sec:proofDegradingOptimal}.


\subsection{Upgrading-merge functions}
The fact that one can merge symbol pairs and get a degraded version of the original channel should come as no surprise. However, it turns out that we can also merge symbol pairs and get an \emph{upgraded} version of the original channel. We first show a simple method of doing this. Later on, we will show a slightly more complex method, and compare between the two.

\begin{figure*}
\hfill
\subfigure[]{\label{subfig:doubleUpgradingWQ}\includegraphics[width=72mm]{{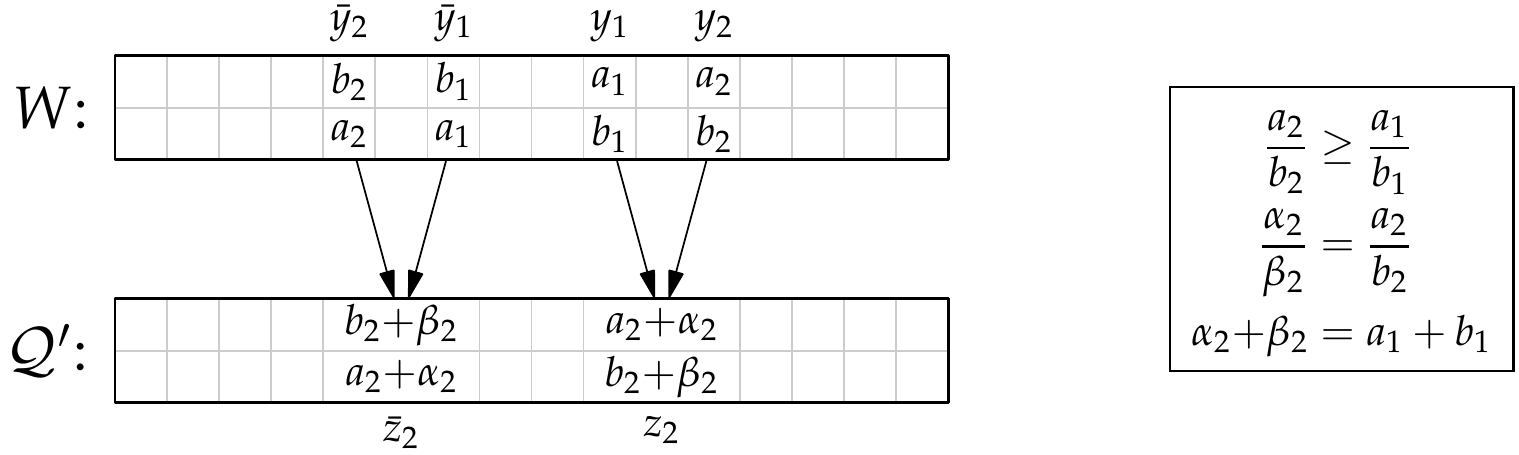}}}
\hfill\hfill
\subfigure[]{\label{subfig:doubleUpgradingP}\includegraphics[width=72mm]{{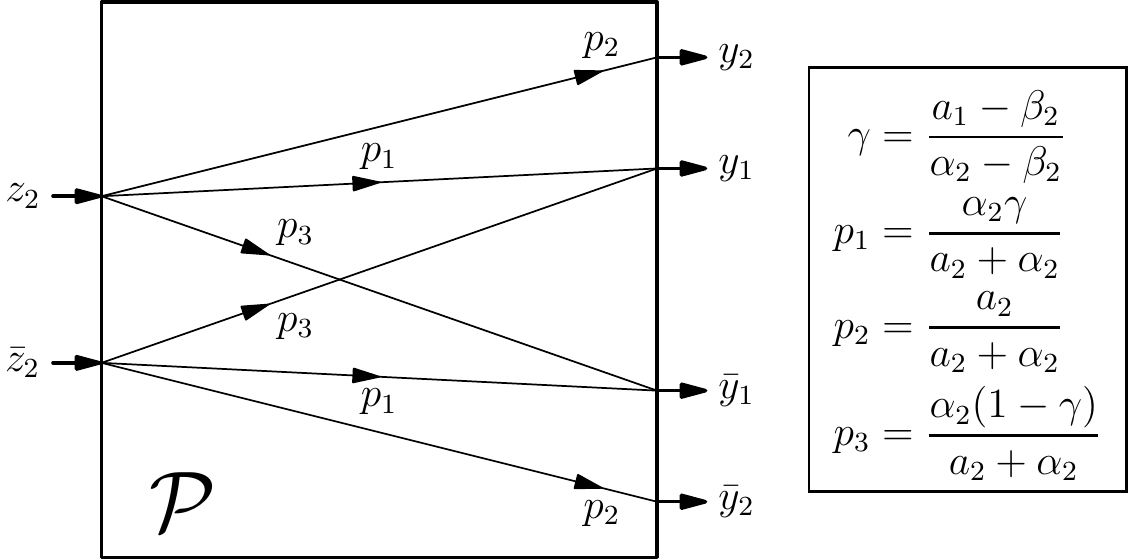}}}
\hfill{}
\caption{First method of Upgrading $\Wgeneric$ to $\Wupgraded$. (a) The upgrading merge operation. (b) The intermediate channel $\intermediateChannel$.}
\label{fig:doubleUpgradingWQP}
\end{figure*}


As in the degrading case, we show how to reduce the output alphabet size by $2$, and then apply this method repeatedly as much as needed. The following lemma shows how the core reduction can be carried out. The intuition behind it is simple. Namely, now we ``promote'' a pair of output symbols to have a higher LR value, and then merge with an existing pair having that LR.

\begin{lemm}
\label{lemm:upgradeStep0}
Let $\Wgeneric:\calX \to \calY$ be a BMS channel, and let $y_2$ and $y_1$ be symbols in the output alphabet $\calY$. Denote $\lambda_2 = \LR(y_2)$ and $\lambda_1 = \LR(y_1)$. Assume that
\begin{equation}
\label{eq:upgradeStep0:lambdaOrder}
1 \leq \lambda_1 \leq \lambda_2 \; .
\end{equation}
Next, let $a_1 = \Wgeneric(y_1|0)$ and $b_1 = \Wgeneric(\bary_1|0)$. Define $\alpha_2$ and $\beta_2$ as follows. If $\lambda_2 < \infty$
\begin{align}
\label{eq:alpha1beta1_step0}
\alpha_2 & = \lambda_2 \frac{a_1 + b_1}{\lambda_2 + 1} &
\beta_2 & = \frac{a_1 + b_1}{\lambda_2 + 1} \; .
\end{align}
Otherwise, we have $\lambda_2 = \infty$, and so define
\begin{align}
\alpha_2 & = a_1 + b_1 &
\beta_2 & = 0 \; .
\end{align}
We note that the subscript ``$2$'' in $\alpha_2$ and $\beta_2$ is meant to suggest a connection to $\lambda_2$, since $\alpha_2/\beta_2=\lambda_2$.

For real numbers $\alpha$, $\beta$, and $x \in \calX$, define
\[
t(\alpha,\beta|x) =
\begin{cases}
\alpha & \mbox{if $x=0$,} \\
\beta & \mbox{if $x=1$.}
\end{cases}
\]
Define the channel $\Wupgraded : \calX \to \calYupgraded$ as follows (see Figure~\ref{subfig:doubleUpgradingWQ}). The output alphabet $\calYupgraded$ is given by
\[
\calYupgraded = \calY \setminus \mysett{y_2,\bary_2,y_1,\bary_1} \cup \mysett{z_2,\barz_2} \; .
\]
For all $x \in \calX$ and $z \in \calYupgraded$,
\[
\Wupgraded(z|x) =
\begin{cases}
\Wgeneric(z|x) & \mbox{if $z \not\in \mysett{z_2,\barz_2}$,} \\
\Wgeneric(y_2|x) + t(\alpha_2,\beta_2|x) & \mbox{if $z = z_2$,} \\
\Wgeneric(\bary_2|x) + t(\beta_2,\alpha_2|x) & \mbox{if $z = \barz_2$.} \\
\end{cases}
\]
Then $\Wupgraded \upgraded \Wgeneric$. That is, $\Wupgraded$ is upgraded with respect to $\Wgeneric$.
\end{lemm}

\begin{proof}
Denote $a_2 = \Wgeneric(y_2|0)$ and $b_2 = \Wgeneric(\bary_2|0)$. First, note that
\[
a_1+b_1 = \alpha_2 + \beta_2 \; .
\]
Next, let $\gamma$ be defined as follows. If $\lambda_2 > 1$, let
\[
\gamma = \frac{a_1-\beta_2}{\alpha_2-\beta_2} = \frac{b_1-\alpha_2}{\beta_2 - \alpha_2} \; ,
\]
and note that (\ref{eq:upgradeStep0:lambdaOrder}) implies that $0 \leq \gamma \leq 1$.
Otherwise ($\lambda_1 = \lambda_2 = 1$), let
\[
\gamma = 1 \; .
\]
Define the intermediate channel $\intermediateChannel : \calYupgraded \to \calY$ as follows. 
\[
\intermediateChannel(y|z) = 
\begin{cases}
1 & \mbox{if $z \not\in\mysett{z_2,\barz_2}$ and $y=z$,}\\
\frac{\alpha_2 \gamma}{a_2+\alpha_2} & \mbox{if $(z,y) \in \mysett{(z_2,y_1),(\barz_2,\bary_1)}$,}\\
\frac{a_2}{a_2+\alpha_2} & \mbox{if $(z,y) \in \mysett{(z_2,y_2),(\barz_2,\bary_2)}$,}\\
\frac{\alpha_2 (1-\gamma)}{a_2+\alpha_2} & \mbox{if $(z,y) \in \mysett{(z_2,\bary_1),(\barz_2,y_1)}$,}\\
0 & \mbox{otherwise.}
\end{cases}
\]
Notice that when $\lambda_2 < \infty$, we have that
\[
\frac{a_2}{a_2+\alpha_2} = \frac{b_2}{b_2+\beta_2} \quad \mbox{and} \quad \frac{\alpha_2}{a_2+\alpha_2} = \frac{\beta_2}{b_2+\beta_2} \; .
\]
Some simple calculations finish the proof.
\end{proof}

The following corollary shows that we do not ``lose anything'' when applying Lemma~\ref{lemm:upgradeStep0} to symbols $y_1$ and $y_2$ such that $\LR(y_1) = \LR(y_2)$. Thus, intuitively, we do not expect to lose much when applying Lemma~\ref{lemm:upgradeStep0} to symbols with ``close'' LR values.

\begin{coro}
\label{coro:equalLRGivesEquivalent}
Let $\Wgeneric$, $\Wupgraded$, $y_1$, and $y_2$ be as in Lemma~\ref{lemm:upgradeStep0}. If $\LR(y_1) = \LR(y_2)$, then $\Wupgraded \equivalent \Wgeneric$. That is, $\Wgeneric$ and $\Wupgraded$ are equivalent. Moreover, all of the above holds if we replace ``Lemma~\ref{lemm:upgradeStep0}'' by ``Lemma~\ref{lemm:degradeStep}''.
\end{coro}
\begin{proof}
The proof follows by noticing that the channel $\Wupgraded$ we get by applying Lemma~\ref{lemm:upgradeStep0} to $\Wgeneric$, $y_1$, and $y_2$, is exactly the same channel we get if we apply Lemma~\ref{lemm:degradeStep} instead. Thus, we have both $\Wupgraded \upgraded \Wgeneric$ and $\Wupgraded \degraded \Wgeneric$.
\end{proof}

In Lemma~\ref{lemm:upgradeStep0}, we have essentially transferred the probability $\Wgeneric(y_1|0) + \Wgeneric(\bary_1|0)$ onto a symbol pair with a higher LR value. We now show a different method of merging that involves dividing the probability $\Wgeneric(y_1|0) + \Wgeneric(\bary_1|0)$ between a symbol pair with higher LR value and a symbol pair with lower LR value. As we will prove latter on, this new approach is generally preferable.

\begin{figure*}
\hfill
\subfigure[]{\label{subfig:tripleUpgradingWQ}\includegraphics[width=72mm]{{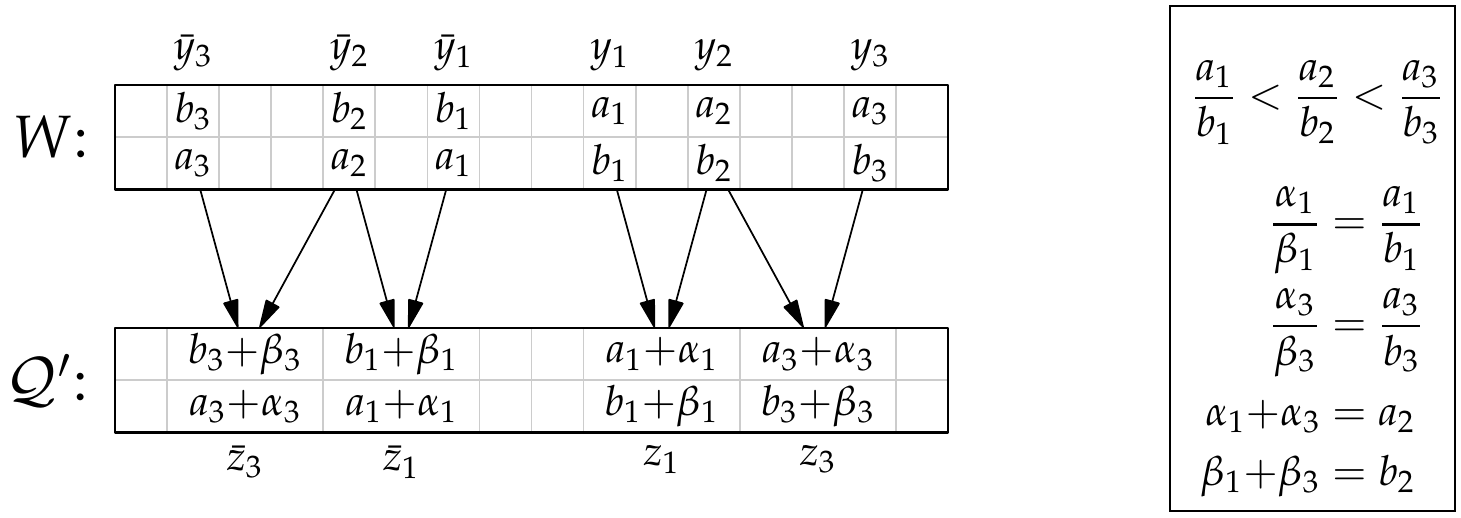}}}
\hfill\hfill
\subfigure[]{\label{subfig:tripleUpgradingP}\includegraphics[width=72mm]{{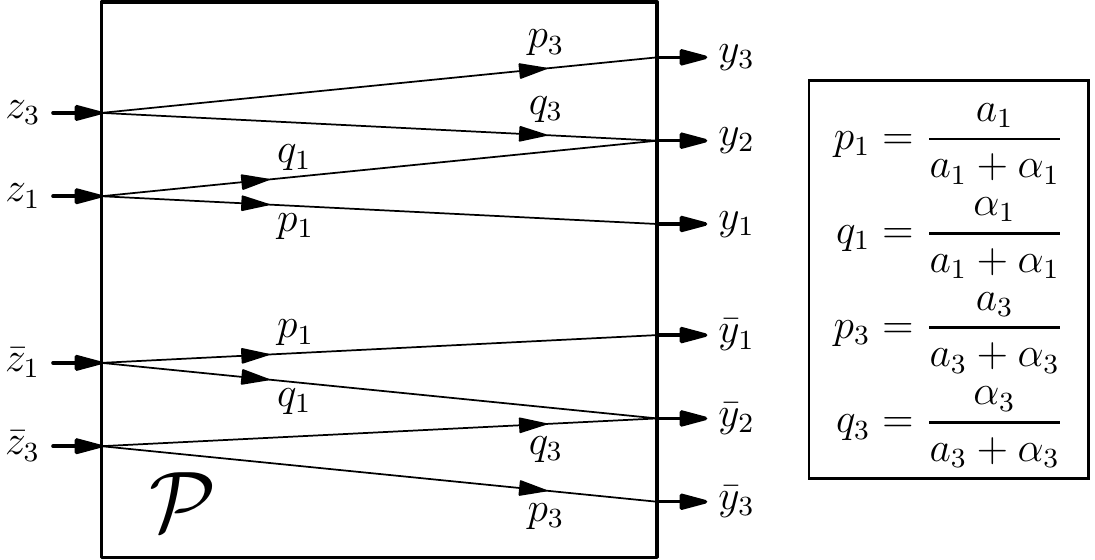}}}
\hfill{}
\caption{Second method of Upgrading $\Wgeneric$ to $\Wupgraded$. (a) The upgrading merge operation. (b) The intermediate channel $\intermediateChannel$.}
\label{fig:tripleUpgradingWQP}
\end{figure*}

\begin{lemm}
\label{lemm:upgradeStep1}
Let $\Wgeneric:\calX \to \calY$ be a BMS channel, and let $y_1$, $y_2$, and $y_3$ be symbols in the output alphabet $\calY$. Denote $\lambda_1 = \LR(y_1)$, $\lambda_2 = \LR(y_2)$, and $\lambda_3 = \LR(y_3)$. Assume that
\[
1 \leq \lambda_1 < \lambda_2 < \lambda_3 \; .
\]
Next, let $a_2 = \Wgeneric(y_2|0)$ and $b_2 = \Wgeneric(\bary_2|0)$. Define $\alpha_1$, $\beta_1$, $\alpha_3$, $\beta_3$ as follows. If $\lambda_3 < \infty$
\begin{align}
\label{eq:alpha1beta1}
\alpha_1 & = \lambda_1 \frac{\lambda_3 b_2 - a_2}{\lambda_3 - \lambda_1} &
\beta_1 & = \frac{\lambda_3 b_2 - a_2}{\lambda_3 - \lambda_1} \; , \\
\label{eq:alpha3beta3}
\alpha_3 & = \lambda_3 \frac{a_2 - \lambda_1 b_2}{\lambda_3 - \lambda_1} &
\beta_3 & = \frac{a_2 - \lambda_1 b_2}{\lambda_3 - \lambda_1} \; .
\end{align}
Otherwise, we have $\lambda_3 = \infty$, and so define
\begin{align}
\alpha_1 & = \lambda_1 b_2 &
\beta_1 & = b_2 \; , \\
\alpha_3 & = a_2 - \lambda_1 b_2 &
\beta_3 & = 0 \; .
\end{align}
Let $t(\alpha,\beta|x)$ be as in Lemma~\ref{lemm:upgradeStep0}, and 
define the \BMS\ channel $\Wupgraded : \calX \to \calYupgraded$ as follows (see Figure~\ref{subfig:tripleUpgradingWQ}). The output alphabet $\calYupgraded$ is given by
\[
\calYupgraded = \calY \setminus \mysett{y_1,\bary_1,y_2,\bary_2,y_3,\bary_3} \cup \mysett{z_1,\barz_1,z_3,\barz_3} \; .
\]
For all $x \in \calX$ and $z \in \calYupgraded$, define
\[
\Wupgraded(z|x) =
\begin{cases}
\Wgeneric(z|x) & \mbox{if $z \not\in \mysett{z_1,\barz_1,z_3,\barz_3}$,} \\
\Wgeneric(y_1|x) + t(\alpha_1,\beta_1|x) & \mbox{if $z = z_1$,} \\
\Wgeneric(\bary_1|x) + t(\beta_1,\alpha_1|x) & \mbox{if $z = \barz_1$,} \\
\Wgeneric(y_3|x) + t(\alpha_3,\beta_3|x) & \mbox{if $z = z_3$,} \\
\Wgeneric(\bary_3|x) + t(\beta_3,\alpha_3|x) & \mbox{if $z = \barz_3$.}
\end{cases}
\]
Then $\Wupgraded \upgraded \Wgeneric$. That is, $\Wupgraded$ is upgraded with respect to $\Wgeneric$.
\end{lemm}

\begin{proof}
Denote $a_1 = \Wgeneric(y_1|0)$, $b_1 = \Wgeneric(\bary_1|0)$, $a_3 = \Wgeneric(y_3|0)$, and $b_3 = \Wgeneric(\bary_3|0)$. Define the intermediate channel $\intermediateChannel : \calYupgraded \to \calY$ as follows. 
\[
\intermediateChannel(y|z) = 
\begin{cases}
1 & \mbox{if $z \not\in\mysett{z_3,\barz_3,z_1,\barz_1}$ and $y=z$,}\\
\frac{a_1}{a_1+\alpha_1} = \frac{b_1}{b_1+\beta_1} & \mbox{if $(z,y) \in \mysett{(z_1,y_1),(\barz_1,\bary_1)}$,}\\
\frac{\alpha_1}{a_1+\alpha_1} = \frac{\beta_1}{b_1+\beta_1} & \mbox{if $(z,y) \in \mysett{(z_1,y_2),(\barz_1,\bary_2)}$,}\\
\frac{a_3}{a_3+\alpha_3} & \mbox{if $(z,y) \in \mysett{(z_3,y_3),(\barz_3,\bary_3)}$,}\\
\frac{\alpha_3}{a_3+\alpha_3} & \mbox{if $(z,y) \in \mysett{(z_3,y_2),(\barz_3,\bary_2)}$,}\\
0 & \mbox{otherwise.}
\end{cases}
\]
Notice that when $\lambda_3 < \infty$, we have that
\[
\frac{a_3}{a_3+\alpha_3} = \frac{b_3}{b_3+\beta_3} \quad \mbox{and} \quad \frac{\alpha_3}{a_3+\alpha_3} = \frac{\beta_3}{b_3+\beta_3} \; .
\]
The proof follows by observing that, whatever the value of $\lambda_3$, 
\[
\alpha_1 + \alpha_3 = a_2 \quad \mbox{and} \quad \beta_1+\beta_3 = b_2 \; .
\]
\end{proof}

The following lemma formalizes why Lemma~\ref{lemm:upgradeStep1} results in a merging operation that is better than that of Lemma~\ref{lemm:upgradeStep0}.

\begin{lemm}
Let $\Wgeneric$, $y_1$, $y_2$, and $y_3$ be as in Lemma~\ref{lemm:upgradeStep1}. Denote by $\WupgradedTriple : \calX \to \calYupgradedTriple$ the result of applying Lemma~\ref{lemm:upgradeStep1} to $\Wgeneric$, $y_1$, $y_2$, and $y_3$. Next, denote by $\WupgradedDoubleShifted : \calX \to \calYupgradedDoubleShifted$ the result of applying Lemma~\ref{lemm:upgradeStep0} to $\Wgeneric$, $y_2$, and $y_3$. Then $\WupgradedDoubleShifted \upgraded \WupgradedTriple \upgraded \Wgeneric$. Namely, in a sense,  $\WupgradedTriple$ is a more faithful representation of $\Wgeneric$ than $\WupgradedDoubleShifted$ is.
\end{lemm}

\begin{proof}
Recall that the two alphabets $\calYupgradedTriple$ and $\calYupgradedDoubleShifted$ satisfy
\begin{align*}
\calYupgradedTriple &= \mysett{z_1,\barz_1,z_3,\barz_3} \cup \calA \; , \\
\calYupgradedDoubleShifted &= \mysett{y_1,\bary_1,z_3,\barz_3} \cup \calA \; ,
\end{align*}
where
\[
\calA = \calY \setminus \mysett{y_1,\bary_1,y_2,\bary_2,y_3,\bary_3}
\]
is the set of symbols not participating in either merge operation.

In order to prove that $\WupgradedTriple$ is degraded with respect to $\WupgradedDoubleShifted$, we must supply a corresponding intermediate channel $\intermediateChannel : \calYupgradedDoubleShifted \to \calYupgradedTriple$. To this end, let 
\[
\lambda_3 = \frac{\WupgradedTriple(z_3|0)}{\WupgradedTriple(z_3|1)} = \frac{\WupgradedDoubleShifted(z_3|0)}{\WupgradedDoubleShifted(z_3|1)} = \frac{\Wgeneric(y_3|0)}{\Wgeneric(y_3|1)}
\]
and
\[
\gamma = \frac{\WupgradedTriple(z_3|0)}{\WupgradedDoubleShifted(z_3|0)} = \frac{\WupgradedTriple(\barz_3|1)}{\WupgradedDoubleShifted(\barz_3|1)} \; .
\]
Note that in both Lemma~\ref{lemm:upgradeStep1} and \ref{lemm:upgradeStep0} we have that $\alpha_3/\beta_3=\lambda_3$. Next, we recall that in Lemma~\ref{lemm:upgradeStep0} we have that $\alpha_3+\beta_3 = a_2+b_2$ whereas in Lemma~\ref{lemm:upgradeStep1} we have $\alpha_3+\beta_3 = a_2+b_2 - \alpha_1 - \beta_1$. Thus, we conclude that $0 \leq \gamma \leq 1$. Moreover, since $0 \leq \gamma \leq 1$, we conclude that the following definition of an intermediate channel is indeed valid.
\begin{multline*}
\intermediateChannel(\zTriple|\zDoubleShifted) = \\
\begin{cases}
1 & \mbox{if $\zTriple = \zDoubleShifted$ and $\zTriple \in \calA$,}\\
1 & \mbox{if $(\zDoubleShifted,\zTriple) \in \mysett{(y_1,z_1),(\bary_1,\barz_1)}$,}\\
\gamma & \mbox{if $(\zDoubleShifted,\zTriple) \in \mysett{(z_3,z_3),(\barz_3,\barz_3)}$,}\\
\frac{(1-\gamma)\lambda_1}{\lambda_1+1} & \mbox{if $(\zDoubleShifted,\zTriple) \in \mysett{(z_3,z_1),(\barz_3,\barz_1)}$,}\\
\frac{(1-\gamma)}{\lambda_1+1} & \mbox{if $(\zDoubleShifted,\zTriple) \in \mysett{(z_3,\barz_1),(\barz_3,z_1)}$,}\\
0 & \mbox{otherwise.}
\end{cases}
\end{multline*}
A short calculation shows that $\intermediateChannel$ is indeed an intermediate channel that degrades $\WupgradedDoubleShifted$ to $\WupgradedTriple$.
\end{proof}

At this point, the reader may be wondering why we have chosen to state Lemma~\ref{lemm:upgradeStep0} at all. Namely, it is clear what disadvantages it has with respect to Lemma~\ref{lemm:upgradeStep1}, but we have yet to indicate any advantages. Recalling the conditions of Lemma~\ref{lemm:upgradeStep1}, we see that it can not be employed when the set $\mysett{\lambda_1,\lambda_2,\lambda_3}$ contains non-unique elements. In fact, more is true. Ultimately, when one wants to implement the algorithms outlined in this paper, one will most probably use floating point numbers. Recall that a major source of numerical instability stems from subtracting two floating point numbers that are too close. By considering the denominator in (\ref{eq:alpha1beta1}) and (\ref{eq:alpha3beta3}) we see that $\lambda_1$ and $\lambda_3$ should not be too close. Moreover, by considering the numerators, we conclude that $\lambda_2$ should not be too close to both $\lambda_1$ and $\lambda_3$. So, when these cases do occur, our only option is Lemma~\ref{lemm:upgradeStep0}.


We now define the merge-upgrading procedure we have used. Apart from an initial step, it is very similar to the merge-degrading procedure we have previously outlined. Assume we are given a BMS channel $\Wgeneric: \calX \to \calY$ with an alphabet size of $2L$ and wish to reduce its alphabet size to $\mu$, while transforming $\Wgeneric$ into a upgraded version of itself. If $2L \leq \mu$, then, as before, we are done. Otherwise, as in the merge-degrading procedure, we choose $L$ representatives $y_1,y_2,\ldots,y_L$, and order them according to their $\LR$ values, all of which are greater than or equal to $1$. We now specify the preliminary step: for some specified parameter epsilon (we have used $\epsilon = 10^{-3}$), we check if there exists an index $1 \leq i < L$ such that the ratio $\LR(y_{i+1}) / \LR(y_i)$ is less than $1+\epsilon$. If so, we apply Lemma~\ref{lemm:upgradeStep0} repeatedly, until no such index exists. Now comes the main step. We ask the following question: for which index $1 \leq i \leq L-1$ does the channel resulting from the application of Lemma~\ref{lemm:upgradeStep1} to $\Wgeneric$, $y_i$, $y_{i+1}$, and $y_{i+2}$ result in a channel with smallest capacity increase? After finding the minimizing index $i$, we indeed apply Lemma~\ref{lemm:upgradeStep1} and get an upgraded channel $\Wupgraded$ with an alphabet size smaller by $2$ than that of $\Wgeneric$. The same process is applied to $\Wupgraded$, until the output alphabet size is not more than $\mu$. As before, assuming the output alphabet size of $\Wgeneric$ is at most $2 \mu^2$, an implementation similar to that given in Algorithm~\ref{alg:degradeChannel} will run in $O(\mu^2 \log \mu)$ time.

As was the case for degrading, the following theorem proves that no generality is lost by only considering merging of consecutive triplets of the form $y_i$, $y_{i+1}$, $y_{i+2}$ in the main step. The proof is given in Appendix~\ref{sec:proofUpgradingOptimal}.

\begin{theo}
\label{theo:upgradeOptimal}
Let $\Wgeneric:\calX \to \calY$ be a BMS channel. Denote by $I_\Wgeneric$ the capacity of $\Wgeneric$ and by $I(y_1,y_2,y_3)$ the capacity one gets by the application of Lemma~\ref{lemm:upgradeStep1} to $\Wgeneric$ and symbols $y_1,y_2,y_3 \in \calY$ such that 
\[
1 \leq \LR(y_1) \leq \LR(y_2) \leq LR(y_3) \; .
\]
Let $\LR(y_1) = \lambda_1$, $\LR(y_2) = \lambda_2$,  $\LR(y_3) = \lambda_3$, $\pi_2 = \Wgeneric(y_2|0) +\Wgeneric(y_2|1)$, and denote the difference in capacities as
\[
\Delta[\lambda_1;\lambda_2,\pi_2;\lambda_3] = I(y_1,y_2,y_3) - I_\Wgeneric \; .
\]
Then, for all $\lambda_1' \leq \lambda_1$ and $\lambda_3' \geq \lambda_3$,
\begin{equation}
\Delta[\lambda_1;\lambda_2,\pi_2;\lambda_3] \leq \Delta[\lambda_1';\lambda_2,\pi_2;\lambda_3'] \; .
\end{equation}
\end{theo}

We end this section by considering the running time of Algorithms~\ref{alg:highLevelDegrade} and \ref{alg:highLevelUpgrade}.
\begin{theo}
\label{theo:totalRunningTime}
Let an underlying BMS channel $\Wunderlying$, a fidelity parameter $\mu$, and codelength $n=2^m$ be given. Assume that the output alphabet size of the underlying channel $\Wunderlying$ is at most $\mu$. The running time of either Algorithm~\ref{alg:highLevelDegrade} or Algorithm~\ref{alg:highLevelUpgrade} is as follows. Approximating a single bit-channel takes $O(m \cdot \mu^2 \log \mu)$ time; approximating all $n$ bit-channels takes $O(n \cdot \mu^2 \log \mu)$ time.
\end{theo}
\begin{proof}
Without loss of generality, we consider Algorithm~\ref{alg:highLevelDegrade}. Recall that the output alphabet size of $\Wunderlying$ is at most $\mu$. Thus, by induction, at the start of each loop the size of the output alphabet of $\QAlg$ is at most $\mu$. Therefore, at each iteration, calculating $\WAlg$ from $\QAlg$ takes $O(\mu^2)$ time, since the output alphabet size of $\WAlg$ is at most $2 \mu^2$. Next, we have seen that each invocation of $\degradingMerge$ takes $O(\mu^2 \log \mu)$ time. The number of times we loop in Algorithm~\ref{alg:highLevelDegrade} is $m$. Thus, for a single bit-channel, the total running time is $O(m \cdot \mu^2 \log \mu)$.

As was explained at the end of Section~\ref{sec:highLevelAlg}, when approximating all $n$ bit channels, the number of distinct channels that need to be approximated is $2n-2$. Thus, the total running time in this case is $O(n \cdot \mu^2 \log \mu)$.
\end{proof}



\section{Channels with Continuous Output Alphabet}
\label{sec:continuousChannels}
Recall that in order to apply either Algorithm~\ref{alg:highLevelDegrade} or \ref{alg:highLevelUpgrade} to an underlying \BMS\ channel $\Wunderlying$, we had to thus far assume that $\Wunderlying$ has a finite output alphabet. In this section, we show two transforms (one degrading and the other upgrading) that transform a \BMS\ channel with a continuous alphabet to a \BMS\ channel with a specified finite output alphabet size. Thus, after applying the degrading (upgrading) transform we will shortly specify to $\Wunderlying$, we will be in a position to apply Algorithm~\ref{alg:highLevelDegrade} (\ref{alg:highLevelUpgrade}) and get a degraded (upgraded) approximation of $\Wbit_i$. Moreover, we prove that both degraded and upgraded versions of our original channels have a guaranteed closeness to the original channel, in terms of difference of capacity.

Let $\Wunderlying$ be a given \BMS\ channel with a continuous alphabet. We will make a few assumptions on $\Wunderlying$. First, we assume that the output alphabet of $\Wunderlying$ is the reals $\reals$. Thus, for $y \in \reals$, let $f(y|0)$ and $f(y|1)$ be the p.d.f.\ functions of the output given that the input was $0$ and $1$, respectively. Next, we require that the symmetry of $\Wunderlying$ manifest itself as
\[
f(y|0) = f(-y|1) \; , \quad \mbox{for all $y \in \reals$} \; .
\]
Also, for notational convenience, we require that
\begin{equation}
\label{eq:fsane}
f(y|0) \geq f(y|1) \; , \quad \mbox{for all $y \geq 0$} \; .
\end{equation}
Note that all of the above holds for the BAWGN channel (after renaming the input $0$ as $-1$).

We now introduce some notation. For $y \geq 0$, define the likelihood ratio of $y$ as
\begin{equation}
\label{eq:lambday}
\lambda(y) = \frac{f(y|0)}{f(y|1)} \; .
\end{equation}
As usual, if $f(y|1)$ is zero while $f(y|0)$ is not, we define $\lambda(y)=\infty$. Also, if both $f(y|0)$ and $f(y|1)$ are zero, then we arbitrarily define $\lambda(y) = 1$. Note that by (\ref{eq:fsane}), we have that $\lambda(y) \geq 1$. 

Under these definitions, a short calculation shows that the capacity of $\Wunderlying$ is 
\[
I(\Wunderlying) = \int_{0}^{\infty} \left( f(y|0) + f(y|1) \right) C[\lambda(y)] \; dy \; ,
\]
where for $1 \leq \lambda < \infty$
\[
C[\lambda] = 1 - \frac{\lambda}{\lambda+1} \log_2 \left( 1 + \frac{1}{\lambda} \right) - \frac{1}{\lambda+1} \log_2 \left( \lambda + 1 \right) \; ,
\]
and (for continuity) we define $C[\infty] = 1$. 

Let $\mu = 2\outputPairsCount$ be the specified size of the degraded/upgraded channel output alphabet. An important property of $C[\lambda]$ is that it is strictly increasing in $\lambda$ for $\lambda \geq 1$. This property is easily proved, and will now be used to show that the following sets form a partition of the non-negative reals. For $1 \leq i \leq \outputPairsCount-1$, let
\begin{equation}
\label{eq:Aigeneral}
A_i = \myset{y \geq 0 : \frac{i-1}{\outputPairsCount} \leq C[\lambda(y)] < \frac{i}{\outputPairsCount} } \; .
\end{equation}
For $i = \outputPairsCount$ we similarly define (changing the second inequality to a weak inequality)
\begin{equation}
\label{eq:Ailast}
A_\outputPairsCount = \myset{y \geq 0 : \frac{\outputPairsCount-1}{\outputPairsCount} \leq C[\lambda(y)] \leq 1 } \; .
\end{equation}
As we will see later on, we must assume that the sets $A_i$ are sufficiently ``nice''. This will indeed be the case of for BAWGN channel.

\subsection{Degrading transform}
\label{subsec:continuousChannels:degrade}
Essentially, our degrading procedure will consist of $\outputPairsCount$ applications of the continuous analog of Lemma~\ref{lemm:degradeStep}. Denote by $\Wdegraded : \calX \to \calZ$ the degraded approximation of $\Wunderlying$ we are going to produce, where
\[
\calZ = \myset{z_1, \barz_1, z_2, \barz_2, \ldots, z_\outputPairsCount, \barz_\outputPairsCount} \; .
\]
We define $\Wdegraded$ as follows.
\begin{align}
\label{Wdegradedzi}
\Wdegraded(z_i|0) = \Wdegraded(\barz_i|1) &= \int_{A_i} f(y|0) \, dy \; , \\
\label{Wdegradedbarzi}
\Wdegraded(\barz_i|0) = \Wdegraded(z_i|1) &= \int_{A_i} f(-y|0) \, dy \; .
\end{align}

\begin{lemm}
The channel $\Wdegraded : \calX \to \calZ$ is a \BMS\ channel such that $\Wdegraded \degraded \Wunderlying$.
\end{lemm}
\begin{proof}
It is readily seen that $\Wdegraded$ is a \BMS\ channel. To prove $\Wdegraded \degraded \Wunderlying$, we now supply intermediate channel $\intermediateChannel : \reals \to \calZ$.
\[
\intermediateChannel(z | y) =
\begin{cases}
1 & \mbox{if $z = z_i$ and $y \in A_i$} \; ,\\
1 & \mbox{if $z = \barz_i$ and $-y \in A_i$} \; ,\\
0 & \mbox{otherwise} \; .
\end{cases}
\]
\end{proof}

The following lemma bounds the loss in capacity incurred by the degrading operation.
\begin{lemm}
The difference in capacities of $\Wdegraded$ and $\Wunderlying$ can be bounded as follows,
\begin{equation}
\label{eq:contDegradingCapDiff}
0 \leq I(\Wunderlying) - I(\Wdegraded) \leq \frac{1}{\outputPairsCount} = \frac{2}{\mu}\; .
\end{equation}
\end{lemm}
\begin{proof}
The first inequality in (\ref{eq:contDegradingCapDiff}) is a consequence of the degrading relation and (\ref{eq:capacityDegrading}). We now turn our attention to the second inequality.

Recall that since the $A_i$ partition the non-negative reals, the capacity of $\Wunderlying$ equals
\begin{equation}
\label{eq:IWunderlyingAsSumOni}
I(\Wunderlying) = \sum_{i=1}^\outputPairsCount \int_{A_i} \left( f(y|0) + f(y|1) \right) C[\lambda(y)] dy \; .
\end{equation}
As for $\Wdegraded$, we start by defining for $1 \leq i \leq \outputPairsCount$ the ratio
\[
\contLR_i = \frac{\Wdegraded(z_i|0)}{\Wdegraded(z_i|1)} \; ,
\]
where the cases of the numerator and/or denominator equaling zero are as in the definition of $\lambda(y)$. By this definition, similarly to the continuous case, the capacity of $\Wdegraded$ is equal to
\begin{equation}
\label{eq:IWdegradedContDef}
I(\Wdegraded) = \sum_{i=1}^\outputPairsCount \left( \Wdegraded(z_i|0) + \Wdegraded(z_i|1) \right) C[\contLR_i] \; .
\end{equation}

Recall that by the definition of $A_i$ in (\ref{eq:Aigeneral}) and (\ref{eq:Ailast}), we have that for all $y \in A_i$,
\[
\frac{i-1}{\outputPairsCount} \leq C[\lambda(y)] \leq \frac{i}{\outputPairsCount} \; .
\]
Thus, by the definition of $\Wdegraded(z_i|0)$ and $\Wdegraded(z_i|1)$ in (\ref{Wdegradedzi}) and (\ref{Wdegradedbarzi}), respectively, we must have by the strict monotonicity of $C$ that
\[
\frac{i-1}{\outputPairsCount} \leq C[\contLR_i] \leq \frac{i}{\outputPairsCount} \; , \quad \mbox{if  $\Wdegraded(z_i|0)>0$} \; .
\]
Thus, for all $y \in A_i$,
\[
\size{C[\contLR_i] - C[\lambda(y)]} \leq \frac{1}{\outputPairsCount} \; , \quad \mbox{if  $\Wdegraded(z_i|0)>0$} \; .
\]
Next, note that $\Wdegraded(z_i|0)>0$ implies $\Wdegraded(z_i|0) + \Wdegraded(z_i|1)>0$. Thus, we may bound $I(\Wdegraded)$ as follows,
\begin{multline*}
I(\Wdegraded) = \sum_{i=1}^\outputPairsCount \left( \Wdegraded(z_i|0) + \Wdegraded(z_i|1) \right) C[\contLR_i] = \\
\sum_{i=1}^\outputPairsCount \int_{A_i} \left( f(y|0) + f(y|1) \right) C[\contLR_i] dy \geq \\
\sum_{i=1}^\outputPairsCount \int_{A_i} \left( f(y|0) + f(y|1) \right) \left( C[\lambda(y)] - \frac{1}{\outputPairsCount} \right) dy = \\
\left( \sum_{i=1}^\outputPairsCount \int_{A_i} \left( f(y|0) + f(y|1) \right) C[\lambda(y)] dy \right) - \frac{1}{\outputPairsCount} = I(\Wunderlying) - \frac{1}{\outputPairsCount} \; ,
\end{multline*}
which proves the second inequality.
\end{proof}

\subsection{Upgrading transform}
In parallel with the degrading case, our upgrading procedure will essentially consist of $\outputPairsCount$ applications of the continuous analog of Lemma~\ref{lemm:upgradeStep0}. Denote by $\Wupgraded : \calX \to \calYupgraded$ the upgraded approximation of $\Wunderlying$ we are going to produce, where
\[
\calYupgraded =  \myset{z_1, \barz_1, z_2, \barz_2, \ldots, z_\outputPairsCount, \barz_\outputPairsCount} \; .
\]
As before, we will show that the loss in capacity due to the upgrading operation is at most $1/\outputPairsCount$.

Let us now redefine the ratio $\contLR_i$. Recalling that the function $C[\lambda]$ is strictly increasing in $\lambda \geq 1$, we deduce that it has an inverse in that range. Thus, for $1 \leq i \leq \outputPairsCount$, we define $\contLR_i \geq 1$ as follows,
\begin{equation}
\label{eq:lambdaiUpgrading}
\contLR_i = C^{-1}\left[\frac{i}{\outputPairsCount}\right] \; .
\end{equation}
Note that for $i = \outputPairsCount$, we have that $\contLR_\outputPairsCount = \infty$. Also, note that for $y \in A_i$ we have by (\ref{eq:Aigeneral}) and (\ref{eq:Ailast}) that
\begin{equation}
\label{eq:lambdayLambdaiupgrading}
1 \leq \lambda(y) \leq \contLR_i \; .
\end{equation}

We now define $\Wupgraded$. For $1 \leq i \leq \outputPairsCount$, let,
\begin{equation}
\label{eq:piiDefinition}
\pi_i = \int_{A_i} \big( f(\alpha|0) + f(-\alpha|0) \big) \, d\alpha \; .
\end{equation}
Then,
\begin{equation}
\label{eq:WupgradedContDef}
\Wupgraded(z|0) =
\begin{cases}
\frac{\contLR_i \pi_i}{\contLR_i+1} & \mbox{if $z=z_i$ and $\contLR_i \neq \infty$} \; , \\
\frac{\pi_i}{\contLR_i+1} & \mbox{if $z=\barz_i$ and $\contLR_i \neq \infty$} \; , \\
\pi_i & \mbox{if $z=z_i$ and $\contLR_i = \infty$} \; , \\
0 & \mbox{if $z=\barz_i$ and $\contLR_i = \infty$} \; , \\
\end{cases}
\end{equation}
and
\begin{equation}
\Wupgraded(z_i|1) = \Wupgraded(\barz_i|0) \; , \quad \Wupgraded(\barz_i|1) = \Wupgraded(z_i|0) \; .
\end{equation}

\begin{lemm}
The channel $\Wupgraded : \calX \to \calYupgraded$ is a \BMS\ channel such that $\Wupgraded \upgraded \Wunderlying$.
\end{lemm}

\begin{proof}
As before, the proof that $\Wupgraded$ is a \BMS\ channel is easy. To show that $\Wupgraded \upgraded \Wunderlying$, we must supply the intermediate channel $\intermediateChannel$. The proof follows easily if we define $\intermediateChannel : \calYupgraded \to \reals$ as the cascade of two channels, $\intermediateChannel_1 : \calYupgraded \to \reals$ and $\intermediateChannel_2 : \reals \to \reals$.

The channel $\intermediateChannel_1 : \calYupgraded \to \reals$ is essentially a renaming channel. Denote by $g(\alpha|z)$ the p.d.f.\ of the output of $\intermediateChannel_1$ given that the input was $z$. Then, for $1 \leq i \leq \outputPairsCount$,
\begin{multline}
g(\alpha|z) = \\
\begin{cases}
\frac{f(\alpha|0)+f(-\alpha|0)}{\pi_i} & \mbox{if $z=z_i$ and $\alpha \in A_i$} \; , \\
\frac{f(\alpha|0)+f(-\alpha|0)}{\pi_i} & \mbox{if $z=\barz_i$ and $-\alpha \in A_i$} \; , \\
0 & \mbox{otherwise} \; .
\end{cases}
\end{multline}
Note that by (\ref{eq:piiDefinition}), the function $g(\alpha|z)$ is indeed a p.d.f.\ for every fixed value of $z \in \calYupgraded$.

Next, we turn to $\intermediateChannel_2 : \reals \to \reals$, the LR reducing channel. Let $\alpha \in A_i$ and recall the definition of $\lambda(y)$ given in
(\ref{eq:lambday}). Define
%
the quantity $p_\alpha$ as follows,
\begin{equation}
p_\alpha = 
\begin{cases}
\frac{\contLR_i - \lambda(\alpha)}{(\lambda(\alpha) + 1)(\contLR_i-1)} & \mbox{if $1 < \contLR_i < \infty$} \; , \\
\frac{1}{2} & \mbox{if $\contLR_i = 1$} \; , \\
\frac{1}{\lambda(\alpha)+1} & \mbox{if $\contLR_i = \infty$ and $\lambda(\alpha) < \infty$} \; , \\
0 & \mbox{if $\lambda(\alpha) = \infty$} \; .
\end{cases}
\end{equation}
By (\ref{eq:lambdayLambdaiupgrading}) with $\alpha$ in place of $y$ we deduce that $0 \leq p_\alpha \leq 1/2$. We define the channel $\intermediateChannel_2 : \reals \to \reals$ as follows. For $y \geq 0$,
\begin{equation}
\intermediateChannel_2(y|\alpha) =
\begin{cases}
1-p_\alpha & \mbox{if $y = \alpha$} \; , \\
p_\alpha & \mbox{if $y = -\alpha$} \; ,
\end{cases}
\end{equation}
and
\[
\intermediateChannel_2(-y|-\alpha) = \intermediateChannel_2(y|\alpha) \; .
\]
Consider the random variable $Y$, which is defined as the output of the concatenation of channels $\Wupgraded$, $\intermediateChannel_1$, and $\intermediateChannel_2$, given that the input to $\Wupgraded$ was $0$. We must show that the p.d.f.\ of $Y$ is $f(y|0)$. To do this, we consider the limit
\[
\lim_{\epsilon \to 0} \frac{\Prob(y \leq Y \leq y+\epsilon)}{\epsilon} \; .
\]
Consider first a $y$ such that $y \in A_i$, and assume further that $\epsilon$ is small enough so that the whole interval between $y$ and $y+\epsilon$ is in $A_i$. In this case, the above can be expanded to
\begin{multline*}
\lim_{\epsilon \to 0} \frac{1}{\epsilon}\Big[ \Wupgraded(z_i | 0) \cdot \int_{y}^{y+\epsilon}g(\alpha|z_i)(1-p_\alpha) \, d\alpha
\\
+ \Wupgraded(\barz_i | 0) \cdot \int_{-y-\epsilon}^{-y}g(\alpha|\barz_i)p_{-\alpha} \, d\alpha \Big] \; .
\end{multline*}
Assuming that the two integrands are indeed integrable, this reduces to
\[
\Wupgraded(z_i | 0) \cdot g(y|z_i)(1-p_y) + \Wupgraded(\barz_i | 0) \cdot g(-y|\barz_i)p_y \; .
\]
From here, simple calculations indeed reduce the above to $f(y|0)$. The other cases are similar.
\end{proof}

As in the degrading case, we can bound the loss in capacity incurred by the upgrading operation.
\begin{lemm}
The difference in capacities of $\Wupgraded$ and $\Wunderlying$ can be bounded as follows,
\begin{equation}
\label{eq:contUpgradingCapDiff}
0 \leq I(\Wupgraded) - I(\Wunderlying) \leq \frac{1}{\outputPairsCount} = \frac{2}{\mu} \; .
\end{equation}
\end{lemm}
\begin{proof}
The first inequality in (\ref{eq:contUpgradingCapDiff}) is a consequence of the upgrading relation and (\ref{eq:capacityDegrading}). We now turn our attention to the second inequality.

For all $y \in A_i$, by (\ref{eq:Aigeneral}), (\ref{eq:Ailast}) and (\ref{eq:lambdaiUpgrading}), we have that
\[
C[\contLR_i] - C[\lambda(y)] \leq \frac{1}{\outputPairsCount} \; , \quad \mbox{if  $\Wupgraded(z_i|0)>0$} \; .
\]
Next, notice that by (\ref{eq:WupgradedContDef}),
\[
\contLR_i = \frac{\Wupgraded(z_i|0)}{\Wupgraded(z_i|1)} \; , \quad \mbox{if $\Wupgraded(z_i|0)>0$} \; .
\]

As in the degrading case, we have that $\Wupgraded(z_i|0)>0$ implies $\Wupgraded(z_i|0) + \Wupgraded(z_i|1)>0$.
Thus, we may bound $I(\Wupgraded)$ as follows,
\begin{multline*}
I(\Wupgraded) = \sum_{i=1}^\outputPairsCount \left( \Wupgraded(z_i|0) + \Wupgraded(z_i|1) \right) C[\contLR_i] = \\
\sum_{i=1}^\outputPairsCount \int_{A_i} \left( f(y|0) + f(y|1) \right) C[\contLR_i] dy \leq \\
\sum_{i=1}^\outputPairsCount \int_{A_i} \left( f(y|0) + f(y|1) \right) \left( C[\lambda(y)] + \frac{1}{\outputPairsCount} \right) dy = \\
\left( \sum_{i=1}^\outputPairsCount \int_{A_i} \left( f(y|0) + f(y|1) \right) C[\lambda(y)] dy \right) + \frac{1}{\outputPairsCount} = I(\Wunderlying) + \frac{1}{\outputPairsCount} \; ,
\end{multline*}
which proves the second inequality.
\end{proof}

\section{Variations of Our Algorithms}
\label{sec:variationsOnTheme}
As one might expect, Algorithms~\ref{alg:highLevelDegrade} and \ref{alg:highLevelUpgrade} can be tweaked and modified. As an example, we now show an improvement to Algorithm~\ref{alg:highLevelDegrade} for a specific case. As we will see in Section~\ref{sec:analysis}, this improvement is key to proving Theorem~\ref{theo:polyConstruction}. Also, it turns out that Algorithm~\ref{alg:highLevelDegrade} is compatible with the result by Guruswami and Xia \cite{GuruswamiXia:13aa}, in the following sense: if we were to use algorithm Algorithm~\ref{alg:highLevelDegrade} with the same $n$ and $\mu$ dictated by \cite{GuruswamiXia:13aa}, then we would be guaranteed a resulting code with parameters as least as good as those promised by \cite{GuruswamiXia:13aa}.


Recall our description of how to construct a polar code given at the end of Section~\ref{sec:highLevelAlg}: obtain a degraded approximation of each bit channel through the use of Algorithm~\ref{alg:highLevelDegrade}, and then select the $k$ best channels when ordered according to the upper bound on the probability of error. Note that Algorithm~\ref{alg:highLevelDegrade} returns a channel, but in this case only one attribute of that channel interests us, namely, the probability of error. In this section, we show how to specialize Algorithm~\ref{alg:highLevelDegrade} accordingly and benefit. 

The specialized algorithm is given as Algorithm~\ref{alg:specialized}. We note that the plots in this paper having to do with an upper bound on the probability of error were produced by running this algorithm.
 The key observation follows from Equations~(26) and (27) in \cite{Arikan:09p}, which we now restate. Recall that $Z(\Wgeneric)$ is the Bhattacharyya parameter of the channel $\Wgeneric$. Then,
\begin{align}
Z(\Wgeneric \ATbad \Wgeneric ) &\leq 2Z(\Wgeneric)-Z(\Wgeneric)^2 \label{eq:ZATbad}\\
Z(\Wgeneric \ATgood \Wgeneric ) &= Z(\Wgeneric)^2 \label{eq:ZATgood}
\end{align}

\LinesNumbered
\begin{algorithm}
\SetInd{0.49em}{0.49em}
\caption{An upper bound on the error probability}
\label{alg:specialized}
\Input{An underlying \BMS\ channel $\Wunderlying$, a bound $\mu=2\outputPairsCount$ on the output alphabet size, a code length $n=2^m$, an index $i$ with binary representation $i=\binaryRep{b_1,b_2,\ldots,b_m}$.}
\Output{An upper bound on $\errorProb(\Wbit_i)$.}
$\ZAlg \leftarrow Z(\Wunderlying)$\;
$\QAlg \leftarrow \degradingMerge(\Wunderlying,\mu)$\; \label{alg:specialized:loopStarts}

\For{$j = 1,2,\ldots, m$}
{
  \eIf{$b_j=0$}
  {
    $\WAlg \leftarrow \QAlg \ATbad \QAlg$ \;
    $\ZAlg \leftarrow \min\{Z(\WAlg), 2\ZAlg-\ZAlg^2 \}$\; \label{alg:specialized:minBad}
  }
  {
    $\WAlg \leftarrow \QAlg \ATgood \QAlg$ \;
    $\ZAlg \leftarrow \ZAlg^2$\; \label{alg:specialized:minGood}
  }
  $\QAlg \leftarrow \degradingMerge(\WAlg,\mu)$\; \label{alg:specialized:degrade}
}
\Return{$\min\{\errorProb(\QAlg), \ZAlg \}$}\; \label{alg:specialized:return}
\end{algorithm}

\begin{theo}
Let a codeword length $n=2^m$, an index $0 \leq i < n$, an underlying channel $\Wunderlying$, and a fidelity parameter $\mu=2\nu$ be given. Denote by $\hat{p}_A$ and $\hat{p}_D$ the outputs of Algorithms~\ref{alg:highLevelDegrade} and \ref{alg:specialized}, respectively. Then,
\[
\hat{p}_A \geq \hat{p}_D \geq \errorProb(\Wbit_i) \; .
\]
That is, the bound produced by Algorithm~\ref{alg:specialized} is always
as least as good as that produced by
Algorithm~\ref{alg:highLevelDegrade}.
\end{theo}

\begin{proof}
Denote by $\Wbit^{(j)}$ the channel we are trying to approximate during iteration $j$. That is, we start with $\Wbit^{(0)} = \Wunderlying$. Then, iteratively $\Wbit^{(j+1)}$ is gotten by transforming $\Wbit^{(j)}$ using either $\ATbad$ or $\ATgood$, according to the value of $b_j$. Ultimately, we have $\Wbit^{(m)}$, which is simply the bit-channel $\Wbit_i$.

The heart of the proof is to show that after iteration $j$ has completed (just after line~\ref{alg:specialized:degrade} has executed), the variable $\ZAlg$ is such that
\[
Z(\Wbit^{(j)}) \leq \ZAlg \leq 1 \; .
\]

The proof is by induction. For the basis, note that before the first iteration starts (just after line~\ref{alg:specialized:loopStarts} has executed), we have $\ZAlg = Z(\Wbit^{(0)})$. For the induction step, first note that $2\ZAlg-\ZAlg^2$ is both an increasing function of $\ZAlg$ and is between $0$ and $1$, when $0 \leq \ZAlg \leq 1$. Obviously, this is also true for $\ZAlg^2$. Now, note that at the end of iteration $j$ we have that the variable $\WAlg$ is degraded with respect to $\Wbit^{(j)}$. Recalling (\ref{eq:BhattacharyyaProbDegrading}), (\ref{eq:ZATbad}) and (\ref{eq:ZATgood}), the induction step is proved.
\end{proof}

\begin{table}
\begin{tabular}{cccc}
& Algorithm~\ref{alg:highLevelDegrade} & Algorithm~\ref{alg:specialized} & Algorithm~\ref{alg:highLevelUpgrade} \\
$\mu=8$ & 5.096030e-03 & 1.139075e-04 & 1.601266e-11 \\
$\mu=16$ & 6.926762e-05 & 2.695836e-05 & 4.296030e-08 \\
$\mu=64$ & 1.808362e-06 & 1.801289e-06 & 7.362648e-07 \\
$\mu=128$ & 1.142843e-06 & 1.142151e-06 & 8.943154e-07 \\
$\mu=256$ & 1.023423e-06 & 1.023423e-06 & 9.382042e-07 \\
$\mu=512$ &  & 9.999497e-07 & 9.417541e-07 \\
\end{tabular}
\caption{Upper and lower bounds on $\PFER$ for $\Wunderlying=\BSC(0.11)$, codeword length $n=2^{20}$, and rate $k/n = 445340/2^{20} = 0.42471$.}
\label{tbl:muandpe}
\end{table}

We end this section by referring to Table~\ref{tbl:muandpe}. In the table, we fix the underlying channel, the codeword length, and the code rate. Then, we compare upper and lower bounds on $\PFER$, for various values of $\mu$. For a given $\mu$, the lower bound is gotten by running Algorithm~\ref{alg:highLevelUpgrade} while the two upper bounds are gotten by running Algorithms \ref{alg:highLevelDegrade} and \ref{alg:specialized}. As can be seen, the upper bound supplied by Algorithm~\ref{alg:specialized} is always superior.

\section{Analysis}
\label{sec:analysis}
As we've seen in previous sections, we can build polar codes by employing Algorithm~\ref{alg:specialized}, and gauge how far we are from the optimal construction by running Algorithm~\ref{alg:highLevelUpgrade}. As can be seen in Figure~\ref{fig:BSCplots}, our construction turns out to be essentially optimal, for moderate sizes of $\mu$. However, we are still to prove Theorem~\ref{theo:polyConstruction}, which gives analytic justification to our method of construction. We do so in this section.

As background to Theorem~\ref{theo:polyConstruction}, recall from \cite{ArikanTelatar:09c} that for a polar code of length $n = 2^m$, the fraction of bit channels with probability of error less than $2^{-n^\beta}$ tends to the capacity of the underlying channel as $n$ goes to infinity, for $\beta < 1/2$. Moreover, the constraint $\beta < 1/2$ is tight in that the fraction of such channels is strictly less than the capacity, for $\beta > 1/2$. Thus, in this context, the restriction on $\beta$ imposed by Theorem~\ref{theo:polyConstruction} cannot be eased.

In order to prove Theorem~\ref{theo:polyConstruction}, we make use of the results of Pedarsani, Hassani, Tal, and Telatar \cite{PHTT:11c}, in particular \cite[Theorem~1]{PHTT:11c} given below. We also point out that many ideas used in the proof of Theorem~\ref{theo:polyConstruction} appear --- in one form or another --- in \cite[Theorem~2]{PHTT:11c} and its proof. 

\begin{theo}[Restatement of {\cite[Theorem~1]{PHTT:11c}}]
\label{theo:CSBound}
Let an underlying BMS channel $\Wunderlying$ be given. Let $n=2^m$ be the code length, and denote by $\Wbit_i^{(m)}$ the corresponding $i$th bit channel, where $0 \leq i < n$. Next, denote by $\Wdegraded_i^{(m)}(\outputPairsCount)$ the degraded approximation of $\Wbit_i^{(m)}$ returned by running Algorithm~\ref{alg:highLevelDegrade} with parameters $\Wunderlying$, $\mu=2\outputPairsCount$, $i$, and $m$. Then,
\[
\frac{\size{\myset{ i : I(\Wbit_i^{(m)}) - I(\Wdegraded_i^{(m)}(\outputPairsCount)) \geq \sqrt{\frac{m}{\outputPairsCount}}}}}{n} \leq \sqrt{\frac{m}{\outputPairsCount}} \; .
\]
\end{theo}

With respect to the above, we remark the following. Recall that in Subsection~\ref{subsec:continuousChannels:degrade} we introduced a method of degrading a continuous channel to a discrete one with at most $\mu=2\outputPairsCount$ symbols. In fact, there is nothing special about the continuous case: a slight modification can be used to degrade an arbitrary discrete channel to a discrete channel with at most $\mu$ symbols. Thus, we have an alternative to the merge-degrading method introduced in Subsection~\ref{subsec:mergingFunctions:degrade}. Thus, it follows easily that Theorem~\ref{theo:CSBound} and thus Theorem~\ref{theo:polyConstruction} would still hold had we used that alternative.

We now break the proof of Theorem~\ref{theo:polyConstruction} into several lemmas. Put simply, the first lemma states that a laxer requirement than that in Theorem~\ref{theo:polyConstruction} on the probability of error  can be met.

\begin{lemm}
\label{lemm:ZdeltaEpsilon}
Let $\Wdegraded_i^{(m)}(\outputPairsCount)$ be as in Theorem~\ref{theo:CSBound}. Then, for every $\delta>0$ and $\epsilon>0$ there exists an $m_0$ and a large enough $\mu=2\outputPairsCount$ such that
\begin{equation}
\label{eq:ZdeltaEpsilon}
\frac{\size{\myset{ i_0 : Z\left(\Wdegraded_{i_0}^{(m_0)}(\outputPairsCount)\right)  \leq \delta}}}{n_0} \geq I(\Wunderlying) - \epsilon \; ,
\end{equation}
where
\[
n_0 = 2^{m_0} \quad \mbox{and} \quad  0 \leq i_0 < n_0 \; .
\]
\end{lemm}

We first note that Lemma~\ref{lemm:ZdeltaEpsilon} has a trivial proof: By \cite[Theorem~2]{Arikan:09p}, we know that there exists an $m_0$ for which (\ref{eq:ZdeltaEpsilon}) holds, if $\Wdegraded_{i_0}^{(m_0)}(\outputPairsCount)$ is replaced by $\Wbit_{i_0}^{(m_0)}$. Thus, we may take $\mu$ large enough so that the pair-merging operation defined in Lemma~\ref{lemm:degradeStep} is never executed, and so $\Wdegraded_{i_0}^{(m_0)}(\outputPairsCount)$ is in fact equal to $\Wbit_{i_0}^{(m_0)}$.

This proof --- although valid --- implies a value of $\mu$ which is
doubly exponential in $m_0$. We now give an alternative proof, which ---
as we have recently learned --- is a precursor to the result of 
Guruswami and Xia~\cite{GuruswamiXia:13aa}. Namely, we state this
alternative proof since we have previously conjectured and now know by
\cite{GuruswamiXia:13aa} that it implies a value of $m_0$ which is not
too large.

\begin{proof}[proof of Lemma~\ref{lemm:ZdeltaEpsilon}]
For simplicity of notation, let us drop the subscript $0$ from $i_0$, $n_0$, and $m_0$. Recall that by \cite[Theorem~1]{Arikan:09p} we have that the capacity of bit channels polarizes. Specifically, for each $\epsilon_1 > 0$ and $\delta_1 >0$ there exists an $m$ such that
\begin{equation}
\label{eq:IpolarizationEpsilonDelta}
\frac{\size{\myset{ i : I\left(\Wbit_{i}^{(m)}\right)  \geq 1 - \delta_1}}}{n} \geq I(\Wunderlying) - \epsilon_1 \; .
\end{equation}
We can now combine the above with Theorem~\ref{theo:CSBound} and deduce that
\begin{multline}
\label{eq:IpolarizationEpsilonDelta_DegradedComplicated}
\frac{\size{\myset{ i : I\left(\Wdegraded_{i}^{(m)}(\outputPairsCount)\right)  \geq 1 - \delta_1-\sqrt{\frac{m}{\outputPairsCount}}}}}{n} \geq \\
I(\Wunderlying) - \epsilon_1 - \sqrt{\frac{m}{\outputPairsCount}} \; .
\end{multline}

Next, we claim that for each $\delta_2 > 0$ and $\epsilon_2 > 0$ there exist $m$ and $\mu = 2\outputPairsCount$ such that
\begin{equation}
\label{eq:eq:IpolarizationEpsilonDelta_DegradedSimple}
\frac{\size{\myset{ i : I\left(\Wdegraded_{i}^{(m)}(\outputPairsCount)\right)  \geq 1 - \delta_2}}}{n} \geq I(\Wunderlying) - \epsilon_2 \; .
\end{equation}
To see this, take $\epsilon_1 = \epsilon_2/2$, $\delta_1 = \delta_2/2$, and let $m$ be the guaranteed constant such that (\ref{eq:IpolarizationEpsilonDelta}) holds. Now, we can take $\outputPairsCount$ big enough so that, in the context of (\ref{eq:IpolarizationEpsilonDelta_DegradedComplicated}), we have that both
\[
\delta_1+\sqrt{\frac{m}{\outputPairsCount}} < \delta_2
\]
and
\[
\epsilon_1 + \sqrt{\frac{m}{\outputPairsCount}} < \epsilon_2 \; .
\]

By \cite[Equation (2)]{Arikan:09p} we have that
\[
Z\left(\Wdegraded_{i}^{(m)}(\outputPairsCount)\right) \leq \sqrt{1-I^2\left(\Wdegraded_{i}^{(m)}(\outputPairsCount)\right)} \; .
\]
Thus, if (\ref{eq:eq:IpolarizationEpsilonDelta_DegradedSimple}) holds then
\[
\frac{\size{\myset{ i : Z\left(\Wdegraded_{i}^{(m)}(\outputPairsCount)\right)  \leq \sqrt{2\delta_2-\delta_2^2}}}}{n} \geq I(\Wunderlying) - \epsilon_2 \; .
\]
So, as before, we deduce that for every $\delta_3 > 0$ and $\epsilon_3 > 0$ there exist $m$ and $\mu$ such that
\[
\frac{\size{\myset{ i : Z\left(\Wdegraded_{i}^{(m)}(\outputPairsCount)\right)  \leq \delta_3}}}{n} \geq I(\Wunderlying) - \epsilon_3 \; .
\]
\end{proof}

The next lemma will be used later to bound the evolution of the variable $\ZAlg$ in Algorithm~\ref{alg:specialized}.
\begin{lemm}
\label{lemm:zetaProcess}
For every $m \geq 0$ and index $0 \leq i < 2^m$ let there be a corresponding real $0 \leq \zeta(i,m) \leq 1$. Denote the binary representation of $i$ by $i=\binaryRep{b_1,b_2,\ldots,b_m}$. Assume that the $\zeta(i,m)$ satisfy the following recursive relation. For $m > 0$ and $i' = \binaryRep{b_1,b_2,\ldots,b_{m-1}}$,
\begin{multline}
\label{eq:zetaCases}
\zeta(i,m) \leq \\
\begin{cases}
2\zeta(i',m-1)-\zeta^2(i',m-1) & \mbox{if $b_m = 0$} \; ,\\
\zeta^2(i',m-1) & \mbox{otherwise} \; .
\end{cases}
\end{multline}

Then, for every $\beta < 1/2$ we have that
\begin{equation}
\label{eq:zetaLimit}
\liminf_{m \to \infty} \frac{\size{\myset{i : \zeta(i,m) < 2^{-n^\beta} }}}{n} \geq 1 - \zeta(0,0) \; ,
\end{equation}
where $n=2^m$.
\end{lemm}

\begin{proof}
First, note that both $f_1(\zeta) = \zeta^2$ and $f_2(\zeta) = 2\zeta - \zeta^2$ strictly increase from $0$ to $1$ when $\zeta$ ranges from $0$ to $1$. Thus, it suffices to prove the claim for the worst case in which the inequality in (\ref{eq:zetaCases}) is replaced by an equality. Assume from now on that this is indeed the case. 

Consider an underlying BEC with probability of erasure (as well as Bhattacharyya parameter) $\zeta(0,0)$. Next, note that the $i$th bit channel, for $0 \leq i < n = 2^m$, is also a BEC, with probability of erasure $\zeta(i,m)$. Since the capacity of the underlying BEC is $1-\zeta(0,0)$, we deduce (\ref{eq:zetaLimit}) by \cite[Theorem~2]{ArikanTelatar:09c}.
\end{proof}

We are now in a position to prove Theorem~\ref{theo:polyConstruction}.
\begin{proof}[Proof of Theorem~\ref{theo:polyConstruction}]
Let us first specify explicitly the code construction algorithm used, and then analyze it. As expected, we simply run Algorithm~\ref{alg:specialized} with parameters $\Wunderlying$ and $n$ to produce upper bounds on the probability of error of all $n$ bit channels. Then, we sort the upper bounds in ascending order. Finally, we produce a generator matrix $G$, with $k$ rows. The rows of $G$ correspond to the first $k$ bit channels according to the sorted order, and $k$ is the largest integer such that the sum of upper bounds is strictly less than $2^{n^{-\beta}}$. By Theorem~\ref{theo:totalRunningTime}, the total running time is indeed $O(n \cdot \mu^2 \log \mu)$.

Recall our definition of $\Wbit_i^{(m)}$ and $\Wdegraded_{i}^{(m)}$ from Theorem~\ref{theo:CSBound}. Denote the upper bound on the probability of error returned by Algorithm~\ref{alg:specialized} for bit channel $i$ by $\errorProbUpperBound(\Wbit_i^{(m)}, \mu)$. The theorem will follow easily once we prove that for all $\epsilon>0$ and $0 < \beta < 1/2$ there exists an even $\mu_0$ such that for all $\mu = 2\outputPairsCount \geq \mu_0$ we have
\begin{equation}
\label{eq:epsilonFromCapacity}
\liminf_{m \to \infty} \frac{\size{\myset{i : \errorProbUpperBound(\Wbit_i^{(m)}, \mu) < 2^{-n^\beta} }}}{n} \geq I(\Wunderlying) - \epsilon \; .
\end{equation}

By Lemma~\ref{lemm:ZdeltaEpsilon}, there exist constants $m_0$ and $\outputPairsCount$ such that
\begin{equation}
\label{eq:ZkappaKappa}
\frac{\size{\myset{ i_0 : Z\left(\Wdegraded_{i_0}^{(m_0)}(\outputPairsCount)\right)  \leq \frac{\epsilon}{2}}}}{n_0} \geq I(\Wunderlying) - \frac{\epsilon}{2} \; ,
\end{equation}
where
\[
n_0 = 2^{m_0} \quad \mbox{and} \quad  0 \leq i_0 < n_0 \; .
\]

Denote the codeword length as $n = 2^m$, where $m = m_0 + m_1$ and $m_1 > 0$. Consider an index $0 \leq i < n$ having binary representation 
\[
i = \binaryRep{b_1,b_2,\ldots,b_{m_0},b_{m_0+1},\ldots,b_m} \; ,
\]
where $b_1$ is the most significant bit. We split the run of Algorithm~\ref{alg:specialized} on $i$ into two stages. The first stage will have $j$ going from $1$ to $m_0$, while the second stage will have $j$ going from $m_0+1$ to $m$.

We start by considering the end of the first stage. Namely, we are at iteration $j=m_0$ and line~\ref{alg:specialized:degrade} has just finished executing. Recall that we denote the value of the variable $\QAlg$ after the line has executed by $\Wdegraded_{i_0}^{(m_0)}(\outputPairsCount)$, where 
\[
i_0 = \binaryRep{b_1,b_2,\ldots,b_{m_0}} \; .
\]
Similarly, define $\ZAlg_{i_0}^{(m_0)}(\outputPairsCount)$ as the value of the variable $\ZAlg$ at that point.
Since, by (\ref{eq:BhattacharyyaProbDegrading}), degrading increases the Bhattacharyya parameter, we have then that the Bhattacharyya parameter of the variable $\WAlg$ is less than or equal to that of the variable $\QAlg$. So, by the minimization carried out in either line \ref{alg:specialized:minBad} or \ref{alg:specialized:minGood}, we conclude the following:
at the end of line~\ref{alg:specialized:degrade} of the algorithm, when $j = m_0$,
\[
\ZAlg = \ZAlg_{i_0}^{(m_0)}(\outputPairsCount)  \leq Z\left(\Wdegraded_{i_0}^{(m_0)}(\outputPairsCount)\right)  = Z(\QAlg) \; .
\]
We can combine this observation with (\ref{eq:ZkappaKappa}) to conclude that
\begin{equation}
\label{eq:ZAlgkappaKappa}
\frac{\size{\myset{ i_0 : \ZAlg_{i_0}^{(m_0)}(\outputPairsCount)  \leq \frac{\epsilon}{2}}}}{n_0} \geq I(\Wunderlying) - \frac{\epsilon}{2} \; .
\end{equation}

We now move on to consider the second stage of the algorithm. Fix an index $i = \binaryRep{b_1,b_2,\ldots,b_{m_0},b_{m_0+1},\ldots,b_m}$. That is, let $i$ have $i_0$ as a binary prefix of length $m_0$. Denote by  $\ZAlg[t]$ the value of $\ZAlg$ at the end of line~\ref{alg:specialized:degrade}, when $j=m_0+t$. By lines \ref{alg:specialized:minBad} and \ref{alg:specialized:minGood} of the algorithm we have, similarly to (\ref{eq:zetaCases}), that

\[
\ZAlg[t+1] \leq
\begin{cases}
2\ZAlg[t]-\ZAlg^2[t] & \mbox{if $b_{m_0+t+1} = 0$} \; ,\\
\ZAlg^2[t] & \mbox{otherwise} \; .
\end{cases}
\]

We now combine our observations about the two stages. Let $\gamma$ be a constant such that
\[
\beta < \gamma < \frac{1}{2} \; .
\]
Considering (\ref{eq:ZAlgkappaKappa}), we see that out of the $n_0=2^{m_0}$ possible prefixes of length $m_0$, the fraction for which
\begin{equation}
\label{eq:Z0good}
\ZAlg[0] \leq \frac{\epsilon}{2} 
\end{equation}
is at least $I(\Wunderlying) - \frac{\epsilon}{2}$. Next, by Lemma~\ref{lemm:zetaProcess}, we see that for each such prefix, the fraction of suffixes for which
\begin{equation}
\label{eq:Zm1}
\ZAlg[m_1] \leq 2^{-(n_1)^\gamma}
\end{equation}
is at least $1-\ZAlg[0]$, as $n_1 = 2^{m_1}$ tends to infinity. 
Thus, for each such prefix, we get by (\ref{eq:Z0good}) that (in the limit) the fraction of such suffixes is at least $1 - \frac{\epsilon}{2}$. We can now put all our bounds together and claim that as $m_1$ tends to infinity, the fraction of indices $0 \leq i < 2^{m}$ for which (\ref{eq:Zm1}) holds is at least
\[
\left(I(\Wunderlying) - \frac{\epsilon}{2}\right) \cdot \left( 1 - \frac{\epsilon}{2} \right) \geq I(\Wunderlying) - \epsilon \; .
\]

By line (\ref{alg:specialized:return}) of Algorithm~\ref{alg:specialized}, we see that $\ZAlg[m_1]$ is an upper bound on the return value $\errorProbUpperBound(\Wbit_i^{(m)})$. Thus, we conclude that
\[
\liminf_{m \to \infty} \frac{\size{\myset{i : \errorProbUpperBound(\Wbit_i^{(m)}, \mu) < 2^{-(n_1)^\gamma} }}}{n} = I(\Wunderlying) - \epsilon
\]
With the above at hand, the only thing left to do in order to prove (\ref{eq:epsilonFromCapacity}) is to show that for $m_1$ large enough we have that
\[
2^{-(n_1)^\gamma} \leq 2^{-n^\beta} \; ,
\]
which reduces to showing that
\[
(n_1)^{\gamma-\beta} \geq (n_0)^\beta \; .
\]
Since $\gamma > \beta$ and $n_0 = 2^{m_0}$ is constant, this is indeed the case.
\end{proof}
We end this section by pointing out a similarity between the analysis used here and the analysis carried out in \cite{HMTU:11a}. In both papers, there are two stages. The first stage (prefix of length $m_0$ in our paper) makes full use of the conditional probability distribution of the channel, while the second stage uses a simpler rule (evolving the bound on the Bhattacharyya parameter in our paper and using an RM rule in \cite{HMTU:11a}). 

\appendices
\section{Proof of Theorem~\ref{theo:degradeOptimal}}
\label{sec:proofDegradingOptimal}
This appendix is devoted to the proof of Theorem~\ref{theo:degradeOptimal}. Although the initial lemmas needed for the proof are rather intuitive, the latter seem to be a lucky coincidence (probably due to a lack of a deeper understanding on the authors' part). The prime example seems to be Equation~(\ref{eq:deltaPrimeSumBingo}) in the proof of Lemma~\ref{lemm:bothDeltaPrimesSmaller}.

We start by defining some notation. Let $\Wgeneric:\calX \to \calY$, $\outputPairsCount$, $y_1,y_2,\ldots,y_\outputPairsCount$ and $\bary_1,\bary_2,\bary_\outputPairsCount$ be as in Theorem~\ref{theo:degradeOptimal}. 
Let $\w \in \calY$ and $\barw \in \calY$ be a symbol pair, and denote by $(a,b)$ the  corresponding probability pair, where
\[
a = p(w | 0 ) = p(\barw | 1) \; , \quad b = p(w | 1 ) = p(\barw | 0) \; .
\]
The contribution of this probability pair to the capacity of $\Wgeneric$ is denoted by
\begin{multline*}
C(a,b) = -(a+b) \log_2((a+b)/2)+a \log_2(a) +b \log_2(b) = \\
-(a+b) \log_2(a+b)+a \log_2(a) +b \log_2(b) + (a+b) \; ,
\end{multline*}
where $0 \log_2 0 = 0$.

Next, suppose we are given two probability pairs: $(a_1,b_1)$ and $(a_2,b_2)$ corresponding to the symbol pair $\w_1,\barw_1$ and $\w_2,\barw_2$, respectively. The capacity difference resulting from the application of Lemma~\ref{lemm:degradeStep} to $\w_1$ and $\w_2$ is denoted by
\[
\Delta(a_1,b_1;a_2,b_2) = C(a_1,b_1) + C(a_2,b_2) - C(a_1+a_2,b_1+b_2) \; .
\]

For reasons that will become apparent later on, we henceforth relax the definition of a probability pair to two non-negative numbers, \emph{the sum of which may be greater than $1$}. Note that $C(a,b)$ is still well defined with respect to this generalization, as is $\Delta(a_1,b_1;a_2,b_2)$. Furthermore, to exclude trivial cases, we require that a probability pair $(a,b)$ has at least one positive element.

The following lemma states that we lose capacity by performing a downgrading merge.
\begin{lemm}
\label{lemm:DeltaNonNegative}
Let $(a_1,b_1)$ and $(a_2,b_2)$ be two probability pairs. Then,
\[
\Delta(a_1,b_1;a_2,b_2) \geq 0 \;
\]
\end{lemm}

\begin{proof}
Assume first that $a_1,b_1,a_2,b_2$ are all positive. In this case, $\Delta(a_1,b_1;a_2,b_2)$ can be written as follows:
\begin{eqnarray*}
(a_1+a_2)\Bigg( &\frac{-a_1}{a_1+a_2} \log_2 \frac{(a_1+b_1)(a_1+a_2)}{a_1(a_1+b_1+a_2+b_2)} + \\ 
& \frac{-a_2}{a_1+a_2} \log_2 \frac{(a_2+b_2)(a_1+a_2)}{a_2(a_1+b_1+a_2+b_2)} \Bigg) + \\
(b_1+b_2)\Bigg( &\frac{-b_1}{b_1+b_2} \log_2 \frac{(a_1+b_1)(b_1+b_2)}{b_1(a_1+b_1+a_2+b_2)} + \\ 
& \frac{-b_2}{b_1+b_2} \log_2 \frac{(a_2+b_2)(b_1+b_2)}{b_2(a_1+b_1+a_2+b_2)} \Bigg)
\end{eqnarray*}
By Jensen's inequality, both the first two lines and the last two lines can be lower bounded be $0$. The proof for cases in which some of the variables equal zero is much the same.
\end{proof}

The intuition behind the following lemma is that the order of merging does matter in terms of total capacity lost.
\begin{lemm}
\label{lemm:DeltaDistributive}
Let $(a_1,b_1)$, $(a_2,b_2)$, and $(a_3,b_3)$ be three probability pairs. Then,
\begin{multline*}
\Delta(a_1,b_1;a_2,b_2) +  \Delta(a_1+a_2,b_1+b_2;a_3,b_3) = \\
\Delta(a_2,b_2;a_3,b_3) +  \Delta(a_1,b_1;a_2+a_3,b_2+b_3)\; .
\end{multline*}
\end{lemm}
\begin{proof}
Both sides of the equation equal
\[
C(a_1,b_1) + C(a_2,b_2) + C(a_3,b_3) - C(a_1+a_2+a_3,b_1+b_2+b_3) \; .
\]

\end{proof}

Instead of working with a probability pair $(a,b)$, we find it easier to work with a probability sum $\pi = a + b$ and likelihood ratio $\lambda = a/b$. Of course, we can go back to our previous representation as follows. If $\lambda = \infty$ then $a = \pi$ and $b=0$. Otherwise, $a = \frac{\lambda \cdot \pi}{\lambda+1}$ and $b = \frac{\pi}{\lambda+1}$. Recall that our relaxation of the term ``probability pair'' implies that $\pi$ is positive and it may be greater than $1$.

Abusing notation, we define the quantity $C$ through $\lambda$ and $\pi$ as well. For $\lambda = \infty$ we have $C[\infty,\pi] = \pi$. Otherwise,
\begin{multline*}
C[\lambda,\pi] = 
\\
\pi \left( 
- \frac{\lambda}{\lambda+1} \log_2 \left( 1+\frac{1}{\lambda} \right) - \frac{1}{\lambda+1} \log_2(1+ \lambda)
\right) + \pi \; .
\end{multline*}

Let us next consider merging operations. The merging of the symbol pair corresponding to $[\lambda_1,\pi_1]$ with that of $[\lambda_2,\pi_2]$ gives a symbol pair with $[\lambda_{1,2},\pi_{1,2}]$, where 
\[
\pi_{1,2} = \pi_1 + \pi_2
\]
and
\begin{multline}
\label{eq:lambdaMean}
\lambda_{1,2} = \bar{\lambda}[\pi_1,\lambda_1;\pi_2, \lambda_2] = \\
\frac{ \lambda_1 \pi_1(\lambda_2+1)+ \lambda_2 \pi_2(\lambda_1+1)}{\pi_1(\lambda_2+1) + \pi_2(\lambda_1+1)} 
\end{multline}

Abusing notation, define
\[
\Delta[\lambda_1,\pi_1;\lambda_2,\pi_2] = C[\lambda_1,\pi_1] + C[\lambda_2,\pi_2] - C[\lambda_{1,2},\pi_{1,2}] \; .
\]
Clearly, we have that the new definition of $\Delta$ is symmetric:
\begin{equation}
\label{eq:DeltaLRSymmetric}
\Delta[\lambda_1,\pi_1;\lambda_2,\pi_2] = \Delta[\lambda_2,\pi_2;\lambda_1,\pi_1] \; .
\end{equation}

\begin{lemm}
\label{lemm:DeltaMonotnicIncreasingInPi}
$\Delta[\lambda_1,\pi_1;\lambda_2,\pi_2]$ is monotonic increasing in both $\pi_1$ and $\pi_2$.
\end{lemm}
\begin{proof}
Recall from (\ref{eq:DeltaLRSymmetric}) that $\Delta$ is symmetric, and so it suffices to prove the claim for $\pi_1$. Thus, our goal is to prove the following for all $\rho > 0$,
\[
\Delta[\lambda_1,\pi_1 + \rho;\lambda_2,\pi_2] \geq \Delta[\lambda_1,\pi_1; \lambda_2,\pi_2] \; .
\]
At this point, we find it useful to convert back from the likelihood ratio/probability sum representation $[\lambda,\pi]$ to the probability pair representation $(a,b)$. Denote by $(a_1,b_1)$, $(a_2,b_2)$, and $(a',b')$ the probability pairs corresponding to $[\lambda_1,\pi_1]$, $[\lambda_2,\pi_2]$, and $[\lambda_1, \pi_1 + \rho]$, respectively. Let $a_3 = a' - a_1$ and $b_3 = b' - b_1$. Next,  since both $(a_1,b_1)$ and $(a',b')$ have the same likelihood ratio, we deduce that both $a_3$ and $b_3$ are non-negative. Under our new notation, we must prove that
\[
\Delta(a_1+a_3,b_1+b_3;a_2,b_2) \geq \Delta(a_1,b_1; a_2,b_2) \; .
\]

Since both $(a_1,b_1)$ and $(a',b')$ have likelihood ratio $\lambda_1$, this is also the case for $(a_3,b_3)$. Thus, a simple calculation shows that
\[
\Delta(a_1,b_1;a_3,b_3) = 0 \; .
\]
Hence,
\begin{multline*}
\Delta(a_1+a_3,b_1+b_3;a_2,b_2) = \\
\Delta(a_1+a_3,b_1+b_3;a_2,b_2) + \Delta(a_1,b_1;a_3,b_3)
\end{multline*}
Next, by Lemma~\ref{lemm:DeltaDistributive},
\begin{multline*}
\Delta(a_1+a_3,b_1+b_3;a_2,b_2) + \Delta(a_1,b_1;a_3,b_3) = \\
\Delta(a_1,b_1;a_2,b_2) + \Delta(a_1+a_2,b_1+b_2;a_3,b_3) \; .
\end{multline*}
Since, by Lemma~\ref{lemm:DeltaNonNegative}, we have that $\Delta(a_1+a_2,b_1+b_2;a_3,b_3)$ is non-negative, we are done.
\end{proof}

We are now at the point in which our relaxation of the term ``probability pair'' can be put to good use. Namely, we will now see how to reduce the number of variables involved by one, by taking a certain probability sum to infinity.

\begin{lemm}
\label{lemm:DeltaLimit}
Let $\lambda_1$, $\pi_1$, and $\lambda_2$ be given. Assume that $0 < \lambda_2 < \infty$. Define
\[
\Delta[\lambda_1,\pi_1;\lambda_2,\infty] = \lim_{\pi_2 \to \infty} \Delta[\lambda_1,\pi_1;\lambda_2,\pi_2] \; .
\]

If $0 < \lambda_1 < \infty$, then
\begin{multline}
\label{eq:DeltaLimit}
\Delta[\lambda_1,\pi_1;\lambda_2,\infty] = \\
\pi_1 \left(
-\frac{\lambda_1}{\lambda_1+1} \log_2\left( \frac{1+\frac{1}{\lambda_1}}{1+\frac{1}{\lambda_2}} \right) - \frac{1}{\lambda_1+1}\log_2\left( \frac{\lambda_1+1}{\lambda_2+1}\right) 
\right) \; .
\end{multline}
If  $\lambda_1 = \infty$, then
\begin{equation}
\label{eq:DeltaLimitLambda1Infinity}
\Delta[\lambda_1,\pi_1;\lambda_2,\infty] = \pi_1 \left(
-\log_2\left( \frac{1}{1+\frac{1}{\lambda_2}} \right) \right) \; .
\end{equation}
If  $\lambda_1 = 0$, then
\begin{equation}
\label{eq:DeltaLimitLambda1Zero}
\Delta[\lambda_1,\pi_1;\lambda_2,\infty] = \pi_1 \left(
- \log_2\left( \frac{1}{\lambda_2+1}\right) 
\right) \; .
\end{equation}
\end{lemm}

\begin{proof}
Consider first the case $0 < \lambda_1 < \infty$. We write out $\Delta[\lambda_1,\pi_1;\lambda_2,\pi_2]$ in full and after rearrangement get
\begin{equation}
\label{eq:DeltaLimBreakupGeneralCase}
\begin{split}
&\pi_1 \left(
\frac{1}{\frac{1}{\lambda_{1,2}}+1} \log_2 \left( 1+ \frac{1}{\lambda_{1,2}} \right)
-\frac{1}{\frac{1}{\lambda_1}+1} \log_2 \left( 1+ \frac{1}{\lambda_1} \right)
\right) + \\
&\pi_1 \left(
\frac{1}{\lambda_{1,2}+1} \log_2 \left( 1+ \lambda_{1,2} \right)
-\frac{1}{\lambda_1+1} \log_2 \left( 1+ \lambda_1 \right)
\right) + \\
&\pi_2 \left(
\frac{1}{\frac{1}{\lambda_{1,2}}+1} \log_2 \left( 1+ \frac{1}{\lambda_{1,2}} \right)
-\frac{1}{\frac{1}{\lambda_2}+1} \log_2 \left( 1+ \frac{1}{\lambda_2} \right)
\right) + \\
&\pi_2 \left(
\frac{1}{\lambda_{1,2}+1} \log_2 \left( 1+ \lambda_{1,2} \right)
-\frac{1}{\lambda_2+1} \log_2 \left( 1+ \lambda_2 \right)
\right) \; ,
\end{split}
\end{equation}
where $\lambda_{1,2}$ is given in (\ref{eq:lambdaMean}). Next, note that 
\[
\lim_{\pi_2 \to \infty} \lambda_{1,2} = \lambda_2 \; .
\]
Thus, applying $\lim_{\pi_2 \to \infty}$ to the first two lines of (\ref{eq:DeltaLimBreakupGeneralCase}) is straightforward. Next, consider the third line of (\ref{eq:DeltaLimBreakupGeneralCase}), and write its limit as
\[
\lim_{\pi_2 \to \infty} \frac{
\frac{1}{\frac{1}{\lambda_{1,2}}+1} \log_2 \left( 1+ \frac{1}{\lambda_{1,2}} \right)
-\frac{1}{\frac{1}{\lambda_2}+1} \log_2 \left( 1+ \frac{1}{\lambda_2} \right)}
{\frac{1}{\pi_2}}
\]
Since $\lim_{\pi_2 \to \infty} \lambda_{1,2} = \lambda_2$, we get that both numerator and denominator tend to $0$ as $\pi_2 \to \infty$. Thus, we apply l'H\^{o}pital's rule and get
\begin{multline*}
\lim_{\pi_2 \to \infty} 
\frac{\frac{1}{(\lambda_{1,2}+1)^2} \left(\log_2 e - \log_2 \left( 1+ \frac{1}{\lambda_{1,2}} \right)\right)\frac{\partial \lambda_{1,2}}{\partial \pi_2}}
{\frac{1}{(\pi_2)^2}} =
\\
\lim_{\pi_2 \to \infty}
\frac{1}{(\lambda_{1,2}+1)^2} \left(\log_2 e - \log_2 \left( 1+ \frac{1}{\lambda_{1,2}} \right)\right) \cdot \\
\frac{\pi_1(\lambda_2+1)(\lambda_1+1)(\lambda_2 - \lambda_1)}{(\frac{\pi_1(\lambda_1+1) + \pi_2(\lambda_1+1)}{\pi_2})^2} = 
\\
\frac{\pi_1(\lambda_2-\lambda_1)}{(\lambda_1+1)(\lambda_2+1)} \left(\log_2 e - \log_2 \left( 1+ \frac{1}{\lambda_2} \right)\right) \; ,
\end{multline*}
where $e=2.71828\ldots$ is Euler's number. Similarly, taking the $\lim_{\pi_2 \to \infty}$ of the fourth line of (\ref{eq:DeltaLimBreakupGeneralCase}) gives
\[
\frac{\pi_1(\lambda_2-\lambda_1)}{(\lambda_1+1)(\lambda_2+1)} \left(-\log_2 e - \log_2 \left( 1+ \lambda_2 \right)\right) \; .
\]
Thus, a short calculations finishes the proof for this case. The cases $\lambda_1 = \infty$ and $\lambda_1 = 0$ are handled much the same way. 
\end{proof}

The utility of the next Lemma is that it asserts a stronger claim than the ``Moreover'' part of Theorem~\ref{theo:degradeOptimal}, for a specific value of $\lambda_2$. 

\begin{lemm}
\label{lemm:bothDeltaPrimesSmaller}
Let probability pairs $(a_1,b_1)$ and $(a_3,b_3)$ have likelihood ratios $\lambda_1$ and $\lambda_3$, respectively. Assume $\lambda_1 \leq \lambda_3$. Denote $\pi_1 = a_1 + b_1$ and $\pi_3 = a_3 + b_3$. Let
\begin{equation}
\label{eq:lambda2_lemm:bothDeltaPrimesSmaller}
\lambda_2 = \lambda_{1,3} = \bar{\lambda}[\pi_1,\lambda_1;\pi_3, \lambda_3]\; ,
\end{equation}
as defined in (\ref{eq:lambdaMean}). Then,
\[
\Delta[\lambda_1,\pi_1;\lambda_2,\infty] \leq \Delta[\lambda_1,\pi_1;\lambda_3,\pi_3]
\]
and
\[
\Delta[\lambda_3,\pi_3;\lambda_2,\infty] \leq \Delta[\lambda_1,\pi_1;\lambda_3,\pi_3]
\]
\end{lemm}

\begin{proof}
We start by taking care of a trivial case. Note that if it is not the case that $0 < \lambda_2 < \infty$, then $\lambda_1 = \lambda_2 = \lambda_3$, and the proof follows easily.

So, we henceforth assume that $0 < \lambda_2 < \infty$, as was done in Lemma~\ref{lemm:DeltaLimit}. Let 
\[
\Delta_{(1,3)} = \Delta[\lambda_1,\pi_1;\lambda_3,\pi_3] \; ,
\]
\[
\Delta'_{(1,2)} = \Delta[\lambda_1,\pi_1;\lambda_2,\infty] \; ,
\]
and
\[
\Delta'_{(2,3)} = \Delta[\lambda_3,\pi_3;\lambda_2,\infty] \; .
\]
Thus, we must prove that $\Delta'_{(1,2)} \leq \Delta_{(1,3)}$ and $\Delta'_{(2,3)} \leq \Delta_{(1,3)}$. Luckily, Lemma~\ref{lemm:DeltaLimit} and a bit of calculation yields that
\begin{equation}
\label{eq:deltaPrimeSumBingo}
\Delta'_{(1,2)} + \Delta'_{(2,3)} = \Delta_{(1,3)} \; .
\end{equation}
Recall that $\Delta'_{(1,2)}$ and $\Delta'_{(2,3)}$ must be non-negative by Lemmas~\ref{lemm:DeltaNonNegative} and \ref{lemm:DeltaMonotnicIncreasingInPi}. Thus, we are done.
\end{proof}

The next lemma shows how to discard the restraint put on $\lambda_2$ in Lemma~\ref{lemm:bothDeltaPrimesSmaller}.
\begin{lemm}
\label{lemm:oneDeltaPrimeSmaller}
Let the likelihood ratios $\lambda_1$, $\lambda_3$ and the probability sums $\pi_1$, $\pi_3$ be as in be as in Lemma~\ref{lemm:bothDeltaPrimesSmaller}. Fix
\begin{equation}
\label{eq:lambda2Range_lemm:bothDeltaPrimesSmaller}
\lambda_1 \leq \lambda_2 \leq \lambda_3 \; .
\end{equation}
Then either
\begin{equation}
\label{eq:leftMergeBetter_lemm:bothDeltaPrimesSmaller}
\Delta[\lambda_1,\pi_1;\lambda_2,\infty] \leq \Delta[\lambda_1,\pi_1;\lambda_3,\pi_3]
\end{equation}
or
\begin{equation}
\label{eq:rightMergeBetter_lemm:bothDeltaPrimesSmaller}
\Delta[\lambda_3,\pi_3;\lambda_2,\infty] \leq \Delta[\lambda_1,\pi_1;\lambda_3,\pi_3]
\end{equation}
\end{lemm}

\begin{proof}
Let $\lambda_{1,3}$ be as in (\ref{eq:lambda2_lemm:bothDeltaPrimesSmaller}), and note that
\[
\lambda_1 \leq \lambda_{1,3} \leq \lambda_3 \; .
\]

Assume w.l.o.g.\ that $\lambda_2$ is such that
\[
\lambda_1 \leq \lambda_2 \leq \lambda_{1,3} \; .
\]
From Lemma~\ref{lemm:bothDeltaPrimesSmaller} we have that
\[
\Delta[\lambda_1,\pi_1;\lambda_{1,3},\infty] \leq \Delta[\lambda_1,\pi_1;\lambda_3,\pi_3]
\]
Thus, we may assume that $\lambda_2 < \lambda_{1,3}$ and aim to prove that
\begin{equation}
\label{eq:lambda2lambda13SubGoal}
\Delta[\lambda_1,\pi_1;\lambda_2,\infty] \leq \Delta[\lambda_1,\pi_1;\lambda_{1,3},\infty] \; .
\end{equation}
Next, notice that 
\begin{equation}
\label{eq:DeltaZeroLambda1Lambda2Equal}
\Delta[\lambda_1,\pi_1;\lambda_2,\infty] = 0 \; , \quad \mbox{if $\lambda_2 = \lambda_1$} \; .
\end{equation}

Thus, let us assume that
\[
\lambda_1 < \lambda_2 < \lambda_{1,3} \; .
\]
Specifically, it follows that
\[
0 < \lambda_2 < \infty
\]
and thus the assumption in Lemma~\ref{lemm:DeltaLimit} holds.

Define the function $f$ as follows
\[
f(\lambda_2') = \Delta[\lambda_1,\pi_1;\lambda_2',\infty] \; .
\]
Assume first that $\lambda_1=0$, and thus by (\ref{eq:DeltaLimitLambda1Zero}) we have that
\[
\frac{\partial f(\lambda_2')}{\partial \lambda_2'} \geq 0 \; .
\]
On the other hand, if $\lambda_1 \neq 0$ we must have that $0 \leq \lambda_1 < \infty$. Thus, by (\ref{eq:DeltaLimit})
we have that
\[
\frac{\partial f(\lambda_2')}{\partial \lambda_2'} = \frac{\pi_1}{(\lambda_1 + 1)(\lambda_2'+1)}\left( 1 - \frac{\lambda_1}{\lambda_2'} \right) \; ,
\]
which is also non-negative for $\lambda_2' \geq \lambda_1$. Thus, we have proved that the derivative is non-negative in both cases, and this together with (\ref{eq:DeltaZeroLambda1Lambda2Equal}) proves (\ref{eq:lambda2lambda13SubGoal}).
\end{proof}

We are now in a position to prove Theorem~\ref{theo:degradeOptimal}.

\begin{proof}[Proof of Theorem \ref{theo:degradeOptimal}]
We first consider the ``Moreover'' part of the theorem. Let $[\lambda_1,\pi_1]$, $[\lambda_2,\pi_2]$, and $[\lambda_1,\pi_1]$ correspond to $y_i$, $y_j$, and $y_k$, respectively. From Lemma~\ref{lemm:oneDeltaPrimeSmaller} we have that either (\ref{eq:leftMergeBetter_lemm:bothDeltaPrimesSmaller}) or (\ref{eq:rightMergeBetter_lemm:bothDeltaPrimesSmaller}) holds. Assume w.l.o.g.\ that (\ref{eq:leftMergeBetter_lemm:bothDeltaPrimesSmaller}) holds. By Lemma~\ref{lemm:DeltaMonotnicIncreasingInPi} we have that
\[
\Delta[\lambda_1,\pi_1; \lambda_2,\pi_2] \leq \Delta[\lambda_1,\pi_1;\lambda_2,\infty] \; .
\]
Thus,
\[
\Delta[\lambda_1,\pi_1; \lambda_2,\pi_2] \leq \Delta[\lambda_1,\pi_1;\lambda_3,\pi_3] \; ,
\]
which is equivalent to
\[
I(y_j,y_k) \geq I(y_i,y_k) \; .
\]

Having finished the ``Moreover'' part, we now turn our attention to the proof of (\ref{eq:theo:degradeOptimal_LRgeq1}). The two equalities in (\ref{eq:theo:degradeOptimal_LRgeq1}) are straightforward, so we are left with proving the inequality. For $\lambda \geq 0$ and $\pi > 0$, the following are easily verified:
\begin{equation}
\label{eq:ClambdaOneOverLambda}
C[\lambda,\pi] = C[1/\lambda,\pi] \; ,
\end{equation}
and
\begin{equation}
\label{eq:CincreasesWithLambda}
\mbox{$C[\lambda,\pi]$ increases with $\lambda \geq 1$} \; .
\end{equation}
Also, for $\bar{\lambda}$ as given in (\ref{eq:lambdaMean}), $\lambda_1,\lambda_2 \geq 0$, and $\pi_1 > 0, \pi_2 > 0$, it is easy to show that
\begin{equation}
\label{eq:lambdaBarIncreasesWithLambdas}
\mbox{$\bar{\lambda}[\lambda_1,\pi_1, \lambda_2,\pi_2]$ increases with both $\lambda_1$ and $\lambda_2$} \; .
\end{equation}

Let $[\lambda_1,\pi_1]$ and  $[\lambda_2,\pi_2]$ correspond to $y_i$ and $y_j$, respectively. Denote
\[
\gamma = \bar{\lambda}[\lambda_1,\pi_1;\lambda_2,\pi_2]
\]
and
\[
\delta = \bar{\lambda}[1/\lambda_1,\pi_1;\lambda_2,\pi_2]
\]
Hence, our task reduces to showing that
\begin{equation}
\label{eq:LR1geq1Reduction}
C[\gamma,\pi_1+\pi_2] \geq C[\delta,\pi_1+\pi_2] \; .
\end{equation}
Assume first that $\delta \geq 1$. Recall that both $\lambda_1 \geq 1$ and $\lambda_2 \geq 1$. Thus, by (\ref{eq:lambdaBarIncreasesWithLambdas}) we conclude that $\gamma \geq \delta \geq 1$. This, together with (\ref{eq:CincreasesWithLambda}) finishes the proof.

Conversely, assume that $\delta \leq 1$. Since
\[
\bar{\lambda}[\lambda_1,\pi_1;\lambda_2,\pi_2] = \bar{\lambda}[1/\lambda_1,\pi_1;1/\lambda_2,\pi_2]
\]
we now get from (\ref{eq:lambdaBarIncreasesWithLambdas}) that $\gamma \leq \delta \leq 1$. This, together with (\ref{eq:ClambdaOneOverLambda}) and (\ref{eq:CincreasesWithLambda}) finishes the proof.

\end{proof}

\section{Proof of Theorem~\ref{theo:upgradeOptimal}}
\label{sec:proofUpgradingOptimal}
As a preliminary step toward the proof of Theorem~\ref{theo:upgradeOptimal}, we convince ourselves that the notation $\Delta[\lambda_1;\lambda_2,\pi_2;\lambda_3]$ used in the theorem is indeed valid. Specifically, the next lemma shows that knowledge of the arguments of $\Delta$ indeed suffices to calculate the difference in capacity. The proof is straightforward.
\begin{lemm}
For $i=1,2,3$, let $y_i$ and $\lambda_i$, as well as $\pi_2$ be as in Theorem~\ref{theo:upgradeOptimal}. If $\lambda_3 < \infty$, then
\begin{multline}
\Delta[\lambda_1;\lambda_2,\pi_2;\lambda_3] = \frac{\pi_2}{(\lambda_2+1)(\lambda_1-\lambda_3)} \Biggl[\\
\shoveright{(\lambda_3-\lambda_2)\left( \lambda_1 \log_2\left( 1 + \frac{1}{\lambda_1} \right) + \log_2(1+\lambda_1) \right) +} \\
\shoveright{(\lambda_2-\lambda_1)\left( \lambda_3 \log_2\left( 1 + \frac{1}{\lambda_3} \right) + \log_2(1+\lambda_3) \right) +} \\
(\lambda_1-\lambda_3)\left( \lambda_2 \log_2\left( 1 + \frac{1}{\lambda_2} \right) + \log_2(1+\lambda_2) \right) 
\Biggr] \; .
\end{multline}
Otherwise, $\lambda_3 = \infty$ and
\begin{multline}
\Delta[\lambda_1;\lambda_2,\pi_2;\lambda_3 = \infty] = \\
\shoveright{\frac{\pi_2}{\lambda_2+1} \Biggl[
-\lambda_1 \log_2\left( 1+ \frac{1}{\lambda_1}\right) - \log_2(1+\lambda_1)} \phantom{\Biggr] \; .}\\
+\lambda_2 \log_2\left( 1+ \frac{1}{\lambda_2}\right) + \log_2(1+\lambda_2) \Biggr] \; .
\end{multline}

\end{lemm}

Having the above calculations at hand, we are in a position to prove Theorem~\ref{theo:upgradeOptimal}.
\begin{proof}[Proof of Theorem~\ref{theo:upgradeOptimal}]
First, let consider the case $\lambda_3 < \infty$. Since our claim does not involve changing the values of $\lambda_2$ and $\pi_2$, let us fix them and denote
\[
f(\lambda_1, \lambda_3) = \Delta[\lambda_1;\lambda_2,\pi_2;\lambda_3] \; .
\]
Under this notation, it suffices to prove that $f(\lambda_1, \lambda_3)$ is decreasing in $\lambda_1$ and increasing in $\lambda_3$, where $\lambda_1 < \lambda_2 < \lambda_3$. A simple calculation shows that
\begin{multline}
\frac{\partial f(\lambda_1, \lambda_3)}{\partial \lambda_1} = \frac{-\pi_2(\lambda_3-\lambda_2)}{(1+\lambda_2)(\lambda_3-\lambda_1)^2}
\Biggl[ \\
\lambda_3 \log \left( \frac{1+\frac{1}{\lambda_1}}{1+\frac{1}{\lambda_3}}\right) +
\log\left(\frac{1+\lambda_1}{1+\lambda_3} \right)
\Biggr] \; .
\end{multline}
So, in order to show that $f(\lambda_1, \lambda_3)$ is decreasing in $\lambda_1$, it suffices to show that the term inside the square brackets is positive for all $\lambda_1 < \lambda_3$. Indeed, if we denote
\[
g(\lambda_1,\lambda_3)  = \lambda_3 \log \left( \frac{1+\frac{1}{\lambda_1}}{1+\frac{1}{\lambda_3}}\right) +
\log\left(\frac{1+\lambda_1}{1+\lambda_3} \right) \; ,
\]
then is readily checked that
\[
g(\lambda_1,\lambda_1) = 0 \; ,
\]
while
\[
\frac{\partial g(\lambda_1,\lambda_3)}{\partial \lambda_1} = \frac{\lambda_3-\lambda_1}{\lambda_1(\lambda_1+1)}
\]
is positive for $\lambda_3 > \lambda_1$. The proof of $f(\lambda_1, \lambda_3)$ increasing in $\lambda_3$ is exactly the same, up to a change of variable names.

Let us now consider the second case, $\lambda_3 = \infty$. Similarly to what was done before, let us fix $\lambda_2$ and $\pi_2$, and consider $\Delta[\lambda_1;\lambda_2,\pi_2;\lambda_3 = \infty]$ as a function of $\lambda_1$. Denote
\[
h(\lambda_1) = \Delta[\lambda_1;\lambda_2,\pi_2;\lambda_3 = \infty] \; .
\]
Under this notation, our aim is to prove that $h(\lambda_1)$ is decreasing in $\lambda_1$. Indeed,
\[
\frac{\partial h(\lambda_1)}{\partial \lambda_1} = \frac{-\pi_2 \log_2\left( 1 + \frac{1}{\lambda_1} \right)}{\lambda_2+1}
\]
is easily seen to be negative.
\end{proof}

\section*{Acknowledgments}
We thank 
Emmanuel Abbe, 
Erdal \Arikan, 
Hamed Hassani,
Ramtin Pedarsani,
Uzi Pereg, 
Eren \c{S}a\c{s}o\u{g}lu, 
Artyom Sharov, 
Emre Telatar, 
and R\"{u}diger Urbanke 
for helpful discussions. 
We are especially grateful to Uzi Pereg and
Artyom~Sharov~for going over many incarnations of this paper and
offering valuable comments.

\twobibs{
\bibliographystyle{IEEEtran}
\bibliography{../../../ee/mybib}

\begin{thebibliography}{10}
\providecommand{\url}[1]{#1}
\csname url@samestyle\endcsname
\providecommand{\newblock}{\relax}
\providecommand{\bibinfo}[2]{#2}
\providecommand{\BIBentrySTDinterwordspacing}{\spaceskip=0pt\relax}
\providecommand{\BIBentryALTinterwordstretchfactor}{4}
\providecommand{\BIBentryALTinterwordspacing}{\spaceskip=\fontdimen2\font plus
\BIBentryALTinterwordstretchfactor\fontdimen3\font minus
  \fontdimen4\font\relax}
\providecommand{\BIBforeignlanguage}[2]{{%
\expandafter\ifx\csname l@#1\endcsname\relax
\typeout{** WARNING: IEEEtran.bst: No hyphenation pattern has been}%
\typeout{** loaded for the language `#1'. Using the pattern for}%
\typeout{** the default language instead.}%
\else
\language=\csname l@#1\endcsname
\fi
#2}}
\providecommand{\BIBdecl}{\relax}
\BIBdecl

\bibitem{Abbe:11a}
E.~Abbe, 
``Extracting randomness and dependencies via a matrix polarization,''
\emph{\textup{\texttt{arXiv:1102.1247v1}}}, 2011.

\bibitem{AbbeTelatar:10a}
E.~Abbe and E.~Telatar, 
``Polar codes for the $m$-user {MAC} and matroids,''
\emph{\textup{\texttt{arXiv:1002.0777v2}}}, 2010.

\bibitem{Arikan:09p}
E.~Ar{\i}kan, 
``Channel polarization: A method for constructing 
capacity-achieving codes for symmetric binary-input memoryless channels,'' 
\emph{IEEE Trans. Inform. Theory}, 
vol.~55, pp.\ 3051--3073, 2009.

\bibitem{Arikan:10a}
E.~Ar{\i}kan, 
``Source polarization,''
\emph{\textup{\texttt{arXiv:1001.3087v2}}}, 2010.

\bibitem{ArikanTelatar:09c}
E.~Ar{\i}kan and E.~Telatar, 
``On the rate of channel polarization,'' 
in \emph{Proc.\ IEEE Symp.\ Inform.\ Theory}, 
Seoul, South Korea, 2009, pp.\ 1493--1495.

\bibitem{BurshteinStrugatski:12a}
D.~Burshtein and A.~Strugatski, 
``Polar write once memory codes,''
\emph{\textup{\texttt{arXiv:1207.0782v2}}}, 2012.

\bibitem{CLRS:01b}
T.H.~Cormen, C.E.~Leiserson, R.L.~Rivest, and C.~Stein, 
\emph{Introduction to Algorithms}, 
2nd~ed. Cambridge, Massachusetts: The MIT Press, 2001.

\bibitem{CoverThomas:06b}
T.M.~Cover and J.A.~Thomas, 
\emph{Elements of Information Theory},
2nd~ed. New York: John Wiley, 2006.

\bibitem{Gallager:68b}
R.G.~Gallager, 
\emph{Information Theory and Reliable Communications}.
New York: John Wiley, 1968.

\bibitem{GoliHassaniUrbanke:12c}
A.~Goli, S.H.~Hassani, and R.~Urbanke, 
``Universal bounds on the scaling behavior of polar codes,''
in \emph{Proc.\ IEEE Symp.\ Inform.\ Theory},  
Cambridge, Massachusetts, 2012, pp.~1957--1961.

\bibitem{GuruswamiXia:13aa}
V.~Guruswami and P.~Xia, 
``Polar codes: Speed of polarization and polynomial gap to capacity,'' 
\url{http://eccc.hpi-web.de/report/2013/050/},
2013.   

\bibitem{HMTU:11a}
S.H.~Hassani, R.~Mori, T.~Tanaka, and R.~Urbanke, 
``Rate-dependent~analysis of the asymptotic 
behavior of channel polarization,''
\emph{\textup{\texttt{arXiv:1110.}}}
\emph{\textup{\texttt{0194v2}}}, 2011.

\bibitem{Korada:09z}
S.B.~Korada, 
``Polar codes for channel and source coding,'' 
Ph.D. dissertation, 
{E}cole {P}olytechnique {F}{\'{e}}d{\'{e}}rale de {L}ausanne,
2009.

\bibitem{KSU:10p}
S.B.~Korada, E.~\c{S}a\c{s}o\u{g}lu, and R.~Urbanke, 
``Polar codes: Characterization of exponent, bounds, and constructions,'' 
\emph{IEEE Trans.\ Inform.\ Theory}, 
vol.~56, pp.\,6253--6264, 2010.

\bibitem{KurkoskiYagi:11a}
B.M.~Kurkoski and H.~Yagi, 
``Quantization of binary-input discrete memoryless
channels with applications to {LDPC} decoding,''
\emph{\textup{\texttt{arXiv:11107.}}}
\emph{\textup{\texttt{5637v1}}}, 2011.

\bibitem{MahdavifarVardy:11p}
H.~Mahdavifar and A.~Vardy, 
``Achieving the secrecy capacity of wiretap channels using polar codes,'' 
\emph{IEEE Trans.\ Inform.\ Theory}, 
vol.~57, pp. 6428--6443, 2011.

\bibitem{Mori:10z}
R.~Mori, 
``Properties and construction of polar codes,'' 
Master's thesis, Kyoto University,
\emph{\textup{\texttt{arXiv:1002.3521}}}, 2010. 

\bibitem{MoriTanaka:09c}
R.~Mori and T.~Tanaka, 
``Performance and construction of polar codes 
on symmetric binary-input memoryless channels,'' 
in \emph{Proc.\ IEEE Symp.\ Inform.\ Theory}, 
Seoul, South Korea, 2009, pp. 1496--1500.

\bibitem{PHTT:11c}
R.~Pedarsani, S.H.~Hassani, I.~Tal, and E.~Telatar, 
``On the construction of polar codes,'' 
in \emph{Proc.\ IEEE Symp.\ Inform.\ Theory}, 
Saint Petersburg, Russia, 2011, pp. 11--15.

\bibitem{RichardsonUrbanke:08b}
T.~Richardson and R.~Urbanke, 
\emph{Modern Coding Theory}.
Cambridge, UK: Cambridge University Press, 2008.

\bibitem{Sasoglu:11c}
E. \c{S}a\c{s}o\u{g}lu, 
``Polarization in the presence of memory,'' 
in \emph{Proc.\ IEEE Symp.\ Inform.\ Theory}, 
Saint Petersburg, Russia, 2011, pp. 189--193.

\bibitem{STA:09a}
E. \c{S}a\c{s}o\u{g}lu, E.~Telatar, and E.~Ar{\i}kan,
``Polarization for arbitrary discrete memoryless channels,''
\emph{\textup{\texttt{arXiv:0908.0302v1}}}, 2009.

\bibitem{STY:10a}
E.~\c{S}a\c{s}o\u{g}lu, E.~Telatar, and E.~Yeh, 
``Polar codes for the two-user multiple-access channel,'' 
\emph{\textup{\texttt{arXiv:1006.4255v1}}}, 2010.

\bibitem{TV:11}
I.~Tal and A.Vardy,
``List decoding of polar codes,''
in \emph{Proc.\ IEEE Symp.\ Inform.\ Theory}, 
Saint Petersburg, Russia, 2011, pp. 1--5.

\end{thebibliography}
}
{
\ifdefined\bibstar\else\newcommand{\bibstar}[1]{}\fi

}

\end{document}